\pdfoutput=1
\documentclass[a4paper,11pt, english]{article}
\usepackage{pool/jheppub}
\usepackage{babel}
\usepackage{graphbox}
\usepackage{multirow}
\usepackage{makecell}
\usepackage{amsmath,amssymb, gensymb}
\usepackage{natbib,hyperref}
\usepackage{url}
\usepackage{siunitx}
\usepackage{xspace}
\usepackage{microtype}
\usepackage{subfig}
\usepackage[useregional]{datetime2}
\usepackage{graphicx}
\usepackage{textcomp}
\usepackage{booktabs}

\usepackage{enumitem}
\usepackage{pgfplots}
\pgfplotsset{compat=1.17}
\usetikzlibrary{patterns}
\usepackage{placeins}
\usepackage{epigraph} 
\usepackage{siunitx}
\usepackage{xspace}

\newcommand{\micro}{\ensuremath{\mu}}

\newcommand{\bbonu}{\ensuremath{\beta\beta0\nu}}
\newcommand{\bbtnu}{\ensuremath{\beta\beta2\nu}}

\newcommand{\tz}{\ensuremath{t_0}}

\newcommand{\stwo}{\ensuremath{S_2}}
\newcommand{\sone}{\ensuremath{S_1}}

\newcommand{\Qbb}{\ensuremath{Q_{\beta\beta}}}

\newcommand{\XEO}{\ensuremath{^{137}}Xe}

\newcommand{\TL}{\ensuremath{{}^{208}\rm{Tl}}}

\newcommand{\BI}{\ensuremath{{}^{214}\rm Bi}}

\newcommand{\xetp}{\ensuremath{{\rm Xe_2^{+}}}}
\newcommand{\nhfp}{\ensuremath{{\rm NH_4^{+}}}}
\newcommand{\nht}{\ensuremath{{\rm NH_3}}}
\newcommand{\TM}{\emph{Topmetal}\xspace}

\newcommand{\Xe}[1]{\ensuremath{^{#1}\mathrm{Xe}}\xspace}

\DeclareSIUnit\c{\mbox{$c$}}
\DeclareSIUnit\magn{\mbox{$\times$}}
\DeclareSIUnit\min{min}
\DeclareSIUnit\week{week}
\DeclareSIUnit\year{yr}
\DeclareSIUnit\years{years}
\DeclareSIUnit\yr{yr}
\DeclareSIUnit\standard{std}
\DeclareSIUnit\str{sr}
\DeclareSIUnit\ppm{ppm}
\DeclareSIUnit\ppb{ppb}
\DeclareSIUnit\ppt{ppt}
\DeclareSIUnit\pe{PE}
\DeclareSIUnit\spe{SPE}
\DeclareSIUnit\ev{events}
\DeclareSIUnit\ct{counts}
\DeclareSIUnit\neutron{\mbox{$n$}}
\DeclareSIUnit\smp{samples}
\DeclareSIUnit\Sample{S}
\DeclareSIUnit\ch{ch}
\DeclareSIUnit\hit{hit}
\DeclareSIUnit\hits{hits}
\DeclareSIUnit\bin{(\mbox{5-PE}~bin)}
\DeclareSIUnit\sgm{\mbox{$\sigma$}}
\DeclareSIUnit\rms{RMS}
\DeclareSIUnit\keVr{\mbox{keV$_{\rm nr}$}}
\DeclareSIUnit\keVee{\mbox{keV$_{e{\rm e}}$}}
\DeclareSIUnit\ph{photon}
\DeclareSIUnit\pes{pes}
\DeclareSIUnit\el{electrons}
\DeclareSIUnit\pm{PMT}
\DeclareSIUnit\inch{"}
\DeclareSIUnit\bit{bit}
\DeclareSIUnit\sample{samples}
\DeclareSIUnit\barn{barn}
\DeclareSIUnit\bara{bar}
\DeclareSIUnit\barg{barg}
\DeclareSIUnit\mlardepth{\mbox(meter~of~\LAr~depth)}
\DeclareSIUnit\Curie{Ci}
\DeclareSIUnit\psi{psi}
\DeclareSIUnit\parsec{pc}
\DeclareSIUnit\liveday{\mbox{live-days}}
\DeclareSIUnit\days{\mbox{days}}
\DeclareSIUnit\day{\mbox{day}}
\DeclareSIUnit\miles{\mbox{miles}}
\DeclareSIUnit\degreeC{\mbox{$^{\circ}$C}}
\DeclareSIUnit\electron{\mbox{$e^-$}}
\DeclareSIUnit\Euro{\mbox{\euro}}
\DeclareSIUnit\cph{cph}
\DeclareSIUnit\neq{neq}
\DeclareSIUnit\unit{unit}
\DeclareSIUnit\byte{Byte}
\DeclareSIUnit\Bq{\becquerel}

\newcommand{\SE}{\ensuremath{e_s}}

\begin{document}

\title{ITACA revisited: Ion Tracking Apparatus with CMOS ASICs}

\author[a,b,1]{J.\,J.~G\'omez-Cadenas,\note{Corresponding authors.}}
\author[c,1]{L.~Arazi,}
\author[e]{G.~Martínez-Lema,}
\author[e]{J.~Renner,}
\author[a,b,1]{S.R.~Soleti,}
\author[a]{S.~Torelli}

\affiliation[a]{Donostia International Physics Center, San Sebasti\'an / Donostia, E-20018, Spain}
\affiliation[b]{Ikerbasque (Basque Foundation for Science), Bilbao, E-48009, Spain}
\affiliation[c]{Ben-Gurion University of the Negev, Beer-Sheva, 8410501, Israel}
\affiliation[e]{Instituto de Física Corpuscular (IFIC, CSIC-UV), 46980, Paterna, Spain}

\emailAdd{jjgomezcadenas@dipc.org}
\emailAdd{larazi@bgu.ac.il}
\emailAdd{roberto.soleti@dipc.org}

\abstract{
High-pressure xenon gas TPCs with electroluminescent amplification (HPXeEL) provide detailed topological reconstruction of charged-particle trajectories, offering a distinctive two-electron signature for neutrinoless double beta decay ($0\beta\beta\nu$) searches. We have recently proposed ITACA, a detector concept that images both the electron track and the corresponding ion track, carried by the positive ions drifting in the opposite direction. While electrons drift rapidly to the anode for standard EL imaging, the positive ions drift slowly to the cathode with millimetre-scale diffusion, allowing time to determine the event energy and barycenter and to position a movable ion detector at the projected arrival point of the ion cloud. We present a conceptual design of the ITACA detector, addressing key feasibility questions. First, we define the detector geometry and operating parameters for a 1-tonne-scale instrument at 15~bar, including a modular tiled electroluminescent structure. Second, we present the conceptual design of the Magnetically Actuated Rotor System (MARS), the mechanism that positions the ion sensor at any $(r, \theta)$ coordinate below the cathode, and show that the expected movement time is fast enough to retain $\sim95\%$ of the drift volume for ion detection, while not significantly perturbing the gas on the scales of the ion drift. Third, we propose using a Topmetal CMOS ASIC-based ion detector as an alternative to the molecular sensor approach described in our original work, enabling real-time, 3D imaging of the ion track without the need for offline laser scanning. Finally, we estimate the sensitivity of the proposed apparatus, showing that enhanced topological discrimination from the ion track, combined with an ultra-low background design, allows exploration of $0\beta\beta\nu$ half-lives in excess of $10^{28}$~yr.
}
\keywords{Neutrinoless double beta decay; TPC; high-pressure xenon chambers; Xenon}
\maketitle
\flushbottom

\section{Introduction}
\label{sec:introduction}

The observation of neutrinoless double-beta decay (\bbonu) would prove that neutrinos are Majorana particles, violate lepton number conservation, and constrain the absolute neutrino mass scale~\cite{Gomez-Cadenas:2023vca}. Searching for this rare transition demands large isotopic masses, ultra-low backgrounds, and excellent energy resolution near the decay Q-value (\Qbb).

The objective of planned next-generation experiments is to achieve half-life sensitivities of $T_{1/2} > 10^{27}$~yr, with an ultimate goal of exceeding $10^{28}$~yr~\cite{Gomez-Cadenas:2023vca}. To attain a sensitivity of $10^{27}$~yr ($10^{28}$~yr), background rates must be reduced to approximately 1 (0.1)~cts/ton/yr within the Region of Interest (ROI), defined as 1~FWHM around \Qbb, requiring total exposures ranging from several to tens of ton-years.

High Pressure Xenon TPCs (HPXe)~\cite{Gomez-Cadenas:2019ges} are one of the leading technologies that can be applied to achieve these ambitious goals. In particular, over the last decade and a half, HPXe TPCs using electroluminescent amplification (HPXeEL) have been developed by the NEXT collaboration, which is currently operating the NEXT-100 apparatus \cite{NEXT:2025yqw} at the Canfranc Underground Laboratory. NEXT-100 can hold $\sim100$~kg of xenon at 15~bar, in a detector of roughly 1~m$^3$ volume; it has already demonstrated an FWHM energy resolution of $(0.93 \pm 0.02)$\% at \Qbb~\cite{NEXT:2025ozn}. 

An HPXe TPC provides a distinctive topological signature that enables identification of events with two emitted electrons. A \bbonu\ candidate is defined as a single, continuous track (with no detached energy deposits), reconstructed with total energy within the ROI around \Qbb\ and fully contained inside the fiducial volume, away from detector surfaces. This ``single-track'' requirement suppresses nearly all backgrounds, leaving only single-electron events with energies near \Qbb, such as those arising from photoelectric interactions (without an accompanying X-ray satellite) of the 2.448~MeV $\gamma$ ray of \BI\ and the 2.615~MeV $\gamma$ ray of \TL\ (the latter, after a first small-angle Compton scattering in the Cu inner shield), and from beta decays of $^{137}$Xe created by cosmogenic activation. Further suppression is achieved by requiring that both track endpoints (``blobs'') carry large energy deposits, a condition satisfied by double beta decay events but not by single-electron backgrounds. The effectiveness of this topological discrimination has been demonstrated by the NEXT collaboration~\cite{Ferrario:2015kta, Ferrario:2019kwg, NEXT:2021vzd, Kekic2021, NEXT:2021pjq}. 

However, the performance of the topological signature in an HPXeEL TPC is limited by diffusion and by electroluminescent blurring. As ionisation electrons drift through the dense gas, they undergo diffusion, producing a charge cloud whose transverse size increases with the drift length. In addition, EL amplification introduces further smearing, as the EL photons are emitted isotropically. Various approaches to reducing diffusion have been investigated, including Xe/He mixtures~\cite{Felkai:2017oeq, NEXT:2019oxh, McDonald_2019}, which may reduce the transverse diffusion coefficient by about a factor of two, and quenching gases~\cite{Gonzalez-Diaz2015, Mistry:2025pew}, which can suppress diffusion to the sub-millimetre level. Alas, quenching gases strongly suppress xenon scintillation, eliminating both the prompt \sone\ signal---essential for event fiducialisation---and the excellent energy resolution afforded by proportional EL amplification. To date, EL-based readout remains the only approach demonstrated to provide simultaneously fiducialization, sub-percent energy resolution, and tracking in large xenon TPCs.

In a recent paper~\cite{Gomez-Cadenas:2025ect}, we proposed a fundamentally different approach to improving the topological signature, which preserves EL amplification. The method relies on recording two complementary tracks for each event: the \emph{electron track}---the established signature in an HPXeEL TPC---and the \emph{ion track}, formed by the positive ions left along the original particle trajectory. The ion track exhibits diffusion at the millimetre level and is free from EL-induced smearing, providing a dramatic boost to topological discrimination. In our original proposal, the ion detector was based on fluorescent molecular sensors, and its use required introducing trace amounts of \nht\ to convert Xe$^+$ ions into \nhfp\ ions that could be trapped by the sensors. We called this concept ITACA (Ion Tracking with Ammonium Cations Apparatus).

In this paper, we present a realistic conceptual design of ITACA, addressing the key feasibility questions left open by our initial proposal. Section~\ref{sec:itaca_det} defines the detector geometry and operating parameters for a large-diameter 1-tonne-scale instrument at 15~bar, including a modular tiled EL structure instead of the standard double-mesh configuration, and \sone/\stwo\ readout with wavelength-shifting fibres mounted on the TPC barrel. Section~\ref{sec:mars} presents the conceptual design of the Magnetically Actuated Rotor System (MARS), the mechanism that positions the ion sensor below the cathode, and shows that the expected movement time is short enough to retain $\sim95\%$ of the drift volume for ion detection. Section~\ref{sec:nausica} introduces the use of a Topmetal CMOS ASIC-based ion detector that enables real-time 3D imaging of the ion cloud, replacing the molecular sensor approach of our original work; the electronic detector permits direct counting of \xetp\ ions without chemical conversion, eliminating the need to add \nht. We retain the name ITACA, now standing for Ion Tracking Apparatus with CMOS ASICs. Section~\ref{sec:sensi} estimates the sensitivity of the proposed apparatus and Section~\ref{sec:conclu} presents our conclusions.

\section{The ITACA Detector}
\label{sec:itaca_det}

\subsection{Design criteria and tradeoffs}

ITACA is an HPXeEL, filled with pure xenon at 15 bar. The pressure is chosen as a tradeoff between mass --- which increases with pressure --- and track topology, which is easier to resolve at lower pressure. Other considerations include ion drift velocity (which decreases linearly with pressure), electron lifetime (which tends to improve at lower pressure), and pressure vessel design. Overall, 15 bar is a good compromise, but operation at 10 or 20 bar is equally feasible. Using electroluminescence to amplify the electron track signal guarantees excellent energy resolution and fiducialization of the event. Pure xenon provides also the primary scintillation needed to establish \tz\ and thus fiducialise the event. Both requirements are essential for a \bbonu\ experiment.
 
 \begin{figure}[!htbp]
    \centering
    \includegraphics[scale=0.25]{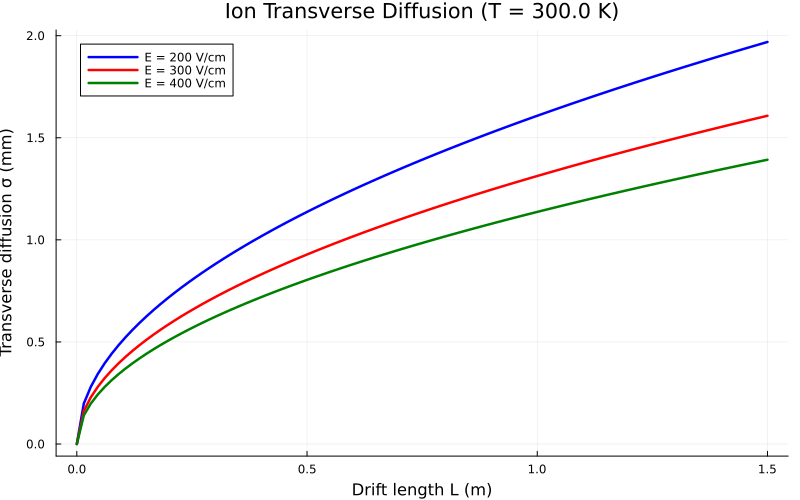}
     \includegraphics[scale=0.25]{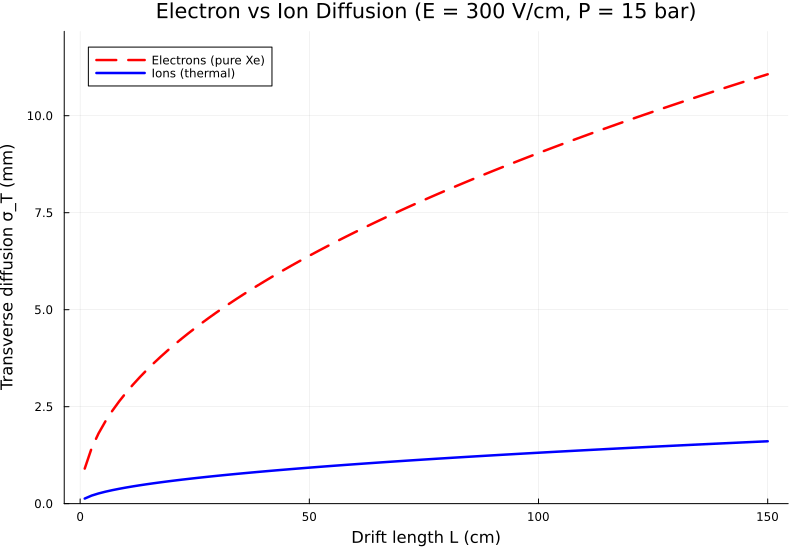}
    \caption{Left: transverse diffusion of \xetp\ ions as a function of drift length for three different drift fields; right: transverse diffusion of \xetp\ ions compared with transverse diffusion of electrons in pure xenon.}
    \label{fig:sigmatvsl}
\end{figure}

\begin{table}[htbp]
\centering
\caption{Summary of ITACA detector parameters.}
\label{tab:itaca_params}
\begin{tabular}{@{}llrl@{}}
\toprule
\textbf{Parameter} & \textbf{Symbol} & \textbf{Value} & \textbf{Unit} \\
\midrule
\multicolumn{4}{@{}l}{\textit{Geometry}} \\
Fiducial diameter      & $D_{\mathrm{fid}}$   & 320   & cm \\
Fiducial height        & $L_{\mathrm{fid}}$   & 150   & cm \\
Fiducial volume        & $V_{\mathrm{fid}}$   & 12.1  & m$^3$ \\
Fiducial Xe mass       &                      & 1050  & kg \\
PV internal diameter   &                      & 3.54  & m \\
PV internal height     &                      & 2.01  & m \\
\midrule
\multicolumn{4}{@{}l}{\textit{Electric field and drift}} \\
Drift field            & $E$                  & 200   & V/cm \\
Electron drift velocity& $v_e$                & 1     & mm/$\mu$s \\
Electron drift time    & $t_e$                & 1.5   & ms \\
Ion drift velocity     & $v_d$                & 10    & cm/s \\
Ion drift time (max)   & $t_{\mathrm{max}}$   & 15    & s \\
\bottomrule
\end{tabular}
\end{table}

 Choosing the detector dimensions implies a tradeoff between the diameter and the length of the TPC. A major design goal is to minimise the transverse diffusion $\sigma_T$ of ions, which in the thermal limit is simply:
 \begin{equation}
\sigma_T =\sqrt{\tfrac{2k_BT}{q_e}\tfrac{L}{E}} 
\label{eq:drift}
\end{equation}
where $L$ is the TPC drift length and $E$ is the drift field. Equation~\ref{eq:drift} favours shorter $L$ and larger $E$. Figure~\ref{fig:sigmatvsl} shows $\sigma_T$ for three different drift fields. Choosing $L = 150$~cm guarantees an average $\sigma_T$ of the order of \SI{1}{mm}, one order of magnitude smaller than the corresponding diffusion for electrons in pure xenon.

Another important design factor is the voltages in the TPC electrodes. All sensors (the CMOS reading the ions and the SiPMs reading the EL light) operate best at ground, but the drifting of electrons and ions requires a constant field in the TPC, and the generation of EL light requires an additional, higher field, in the EL region. 

As we will see in section \ref{sec:mars},  the cathode is instrumented with an Ion Focusing Grid (IFG), needed to block the gas disturbances introduced by the MARS system and to focus the  incoming ions into the reading pads of the CMOS detector. The upper electrode of the IFC is at a voltage, $V_c = 1100$~V. The voltage on the anode side of the TPC depends on the drift length and the drift field. A good compromise between minimising diffusion and keeping voltage at the anode reasonable is to choose a drift field of \SI{200}{V/cm}, which results in \SI{31.1}{kV} at the first anode electrode, and \SI{47.1}{kV} at the second (with \SI{16}{kV} across the FAT-GEM). Achieving these voltages appears reasonable, given the experience with the NEXT experiment~\cite{NEXT:2025yqw}.
 
 Once the length of the TPC is defined, the diameter of the TPC defines the fiducial mass that can be achieved. At 15 bar, choosing $D= 320$ cm, results in a fiducial mass slightly above one ton, $M_{fid} = 1050$ kg. The large diameter, however, has several implications. 
 
 First, it requires designing a pressure vessel (PV) rigid enough to hold 15 bar at this large diameter, while also offering good radiopurity. Titanium-Gr5 is a good choice\footnote{Titanium is intrinsically radiopure, but grade 5 is an alloy which can jeopardise radiopurity, unless a strict selection is carried out. As a backup one could consider 316-Ti steel, used by the PV of the NEXT experiment.}.
 
 Second, the instrumentation of the anode (the silicon plane needed to measure the energy and reconstruct the initial topology of the track, and in particular its barycenter) increases with $R_{TPC}^2$, as do the electronic channels needed to read out the SiPMs. This is, however, a manageable problem, as we will discuss below, given the (relative) sparse instrumentation needed, the low cost of SiPMs and the likely availability of low-cost ASICs for the SiPMs. 
 
 However, the major stumbling block for a large-diameter TPC able to read the ion track is the instrumentation of the cathode. Previous work involving the detection of
 ions for \bbonu\ experiments \cite{Nygren_2018, mei2020topmetalcmosdirectcharge} have quantified the daunting challenge of instrumenting a large surface with CMOS detectors. Instead, the solution proposed here is, within limits, almost independent of the TPC diameter. It involves a Magnetically Actuated Rotor System (MARS), able to move the target to the landing point of the ions in the cathode. Quite literally, in our proposal Muhammad goes to the mountain instead of the other way around. 

This is only possible thanks to the slow velocity of ions with respect to electrons, which provides the time necessary to:

\begin{enumerate}[label=\alph*)]
    \item compute the energy of the event and decide if the event is interesting or not --- the excellent energy resolution afforded by an EL TPC allows selecting only events in a ROI of around 25~keV around \Qbb, thus reducing the rate to a tiny fraction of the total interactions in the detector;
    \item reconstruct the event topology using the silicon plane, further restricting the interesting events to those with a single track in the fiducial volume, with no additional energy deposits;
    \item compute the barycenter of the event, which predicts the transverse coordinates of the ion track at the cathode;
    \item use \tz\ to predict the time at which ions will reach the cathode.
\end{enumerate}

Table~\ref{tab:itaca_params} shows the main parameters of a 1-ton ITACA detector. With some optimisations and tradeoffs, a module in the range of 2--3~ton mass is possible. Larger drift distances are possible tolerating a modest increase in the ion diffusion (to recall, $\sigma_T \propto \sqrt{L}$), the MARS design can be extended to larger diameter, and the pressure increased to $\sim 20$~bar. However, a module with a fiducial mass of 1~ton represents an order-of-magnitude scale over the current state-of-the-art (the NEXT-100 detector) and represents a reasonable compromise between mass and performance. 

\subsection{Main subsystems}

\begin{figure}[!htbp]
    \centering
    \includegraphics[scale=0.68]{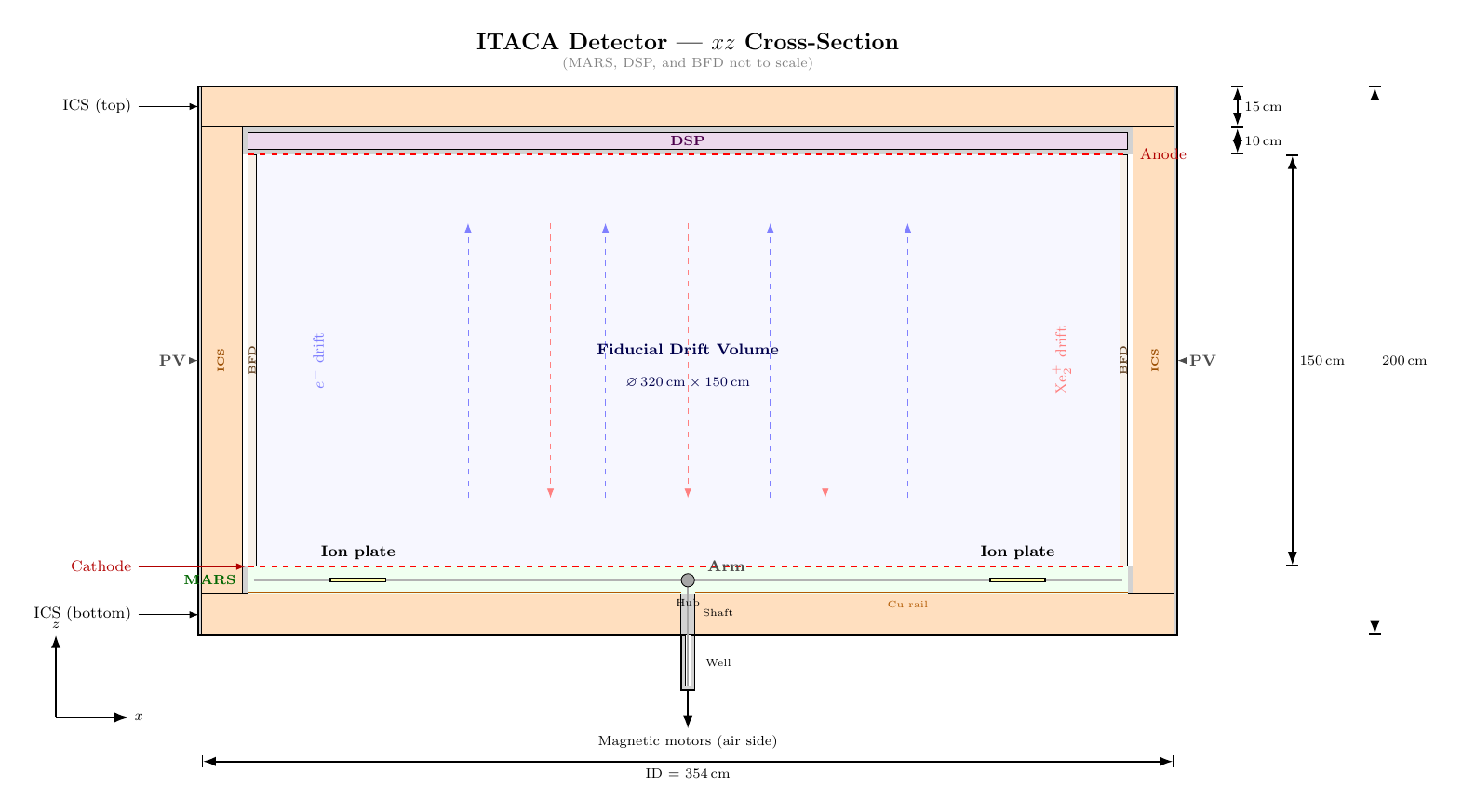}
    \caption{Cross-section (XZ view) of the ITACA detector showing the main subsystems. Not to scale.}
    \label{fig:itaca}
\end{figure}

\begin{table}[htbp]
\centering
\caption{Vertical detector stack (anode at top).}
\label{tab:stack}
\begin{tabular}{@{}llS[table-format=3.1]@{}}
\toprule
\textbf{Component} & \textbf{Description} & {\textbf{Height (cm)}} \\
\midrule
Inner Copper Shield (top)   & OFHC Cu, radiopurity shield           & 15.0  \\
Dense Silicon Plane (DSP)   & SiPM tracking plane on Kapton         & 0.5   \\
Light Guide Honeycomb (LGH) & PMMA rods coupled to SiPMs            & 9.5   \\
FAT-GEM                     & EL amplification structure            & 0.5   \\
Fiducial drift volume       & Active detection region               & 150.0 \\
MARS region        & Arm + ion plate + clearances & 20.0 \\
Inner Copper Shield (bottom)& OFHC Cu, with central shaft bore      & 15.0  \\
\midrule
\textbf{Total internal height} &                                    & \textbf{211.0} \\
\bottomrule
\end{tabular}
\end{table}

Figure~\ref{fig:itaca} shows the apparatus, and 
Table~\ref{tab:stack} summarises the vertical arrangement of detector components from anode to the bottom copper shield. The detector is housed in a cylindrical Ti~Gr.5 pressure vessel. The vessel is lined on all internal surfaces with a \SI{15}{cm} thick Inner Copper Shield (ICS) of ultra-pure OFHC copper, which provides the primary shield against radiogenic backgrounds (including the PV itself).

The ion readout requires a positioning system --- the Magnetically Actuated Rotor System (MARS) --- to place a small CMOS sensor ($160 \times 160$~mm$^2$) at the predicted arrival point of each ion cluster before it reaches the cathode. MARS is described in section~\ref{sec:mars}. The rest of the detector components are described below.

\subsection{TPC and EL Detection System}
\label{sec:tpc}

The Time Projection Chamber constitutes the active detection volume of ITACA. Ionising events produce primary electron--ion pairs in the xenon gas. Primary ionisation electrons drift upward under a uniform electric field and enter a EL modular amplification structure (ELMAS) where they produce secondary scintillation photons (\stwo). About half of these photons are detected by a Dense Silicon Plane (DSP), a large array of SiPMs which reconstructs the electron track and measures its energy and arrival time.  A fraction of the light moving backwards is detected by the Barrel Fibre Detector (BFD), which is also able to measure primary scintillation \sone\ (and thus the start-of-the-event $t_0$). The combination of the DSP and BFD measurements yields an energy resolution at the sub-percent level~\cite{NEXT:2025yqw, NEXT:2020amj}.

\subsubsection*{Drift field and voltage scheme}

The voltage scheme is referenced to the ion detector, which sits at ground potential below the ion focusing grid (IFG) which defines the cathode and is discussed in detail in section \ref{sec:mars}. The top electrode of the IFG is biased at $+\SI{1100}{V}$. From the cathode upward, the voltage increases linearly along the field cage at \SI{200}{V/cm} over the \SI{150}{cm} drift length, reaching $+\SI{31.1}{kV}$ at the entry face of the FAT-GEM. The FAT-GEM adds \SI{16}{kV} across its \SI{5}{mm} gap, so its exit face sits at $+\SI{47.1}{kV}$. This voltage is then degraded to ground across the \SI{9.5}{cm} Light Guide Honeycomb, where the DSP SiPMs operate at ground potential. 

Experience operating the NEXT-100 detector suggests that a drift field of \SI{200}{V/cm} is sufficient to achieve high electron lifetime once the gas is sufficiently pure. At this field, electrons drift to the anode at a velocity of approximately \SI{1}{mm/\micro s}, while the ion drift velocity is $v_d \approx \SI{10}{cm/s}$. The maximum ion drift time, for events originating near the anode, is therefore $t_{\mathrm{max}} = 150/10 = \SI{15}{s}$. The slow drift of the ions has the advantage of improving the fiducial region, as it allows more time for the MARS positioning cycle (see section~\ref{sec:mars}).

\subsubsection*{Field Cage}

The fiducial drift volume is a cylinder of diameter \SI{320}{cm} and height $L_{\mathrm{fid}} = \SI{150}{cm}$, defined by the field cage rings on the sides, the Electroluminescence Modular Amplification Structure at the top, and the Ion Focusing Grid at the bottom.


The field cage is composed of OFHC Cu rings connected by custom-made ultra-low background resistors~\cite{NEXT:2025yqw}. Its role is to maintain a uniform drift field of \SI{200}{V/cm} between the cathode and the FAT-GEM entry, over a total potential difference of \SI{30}{kV}.

\subsubsection*{Electroluminescence Modular Amplification Structure}

\begin{figure}[!htbp]
    \centering
  \includegraphics[width=15cm, height=8cm, keepaspectratio]{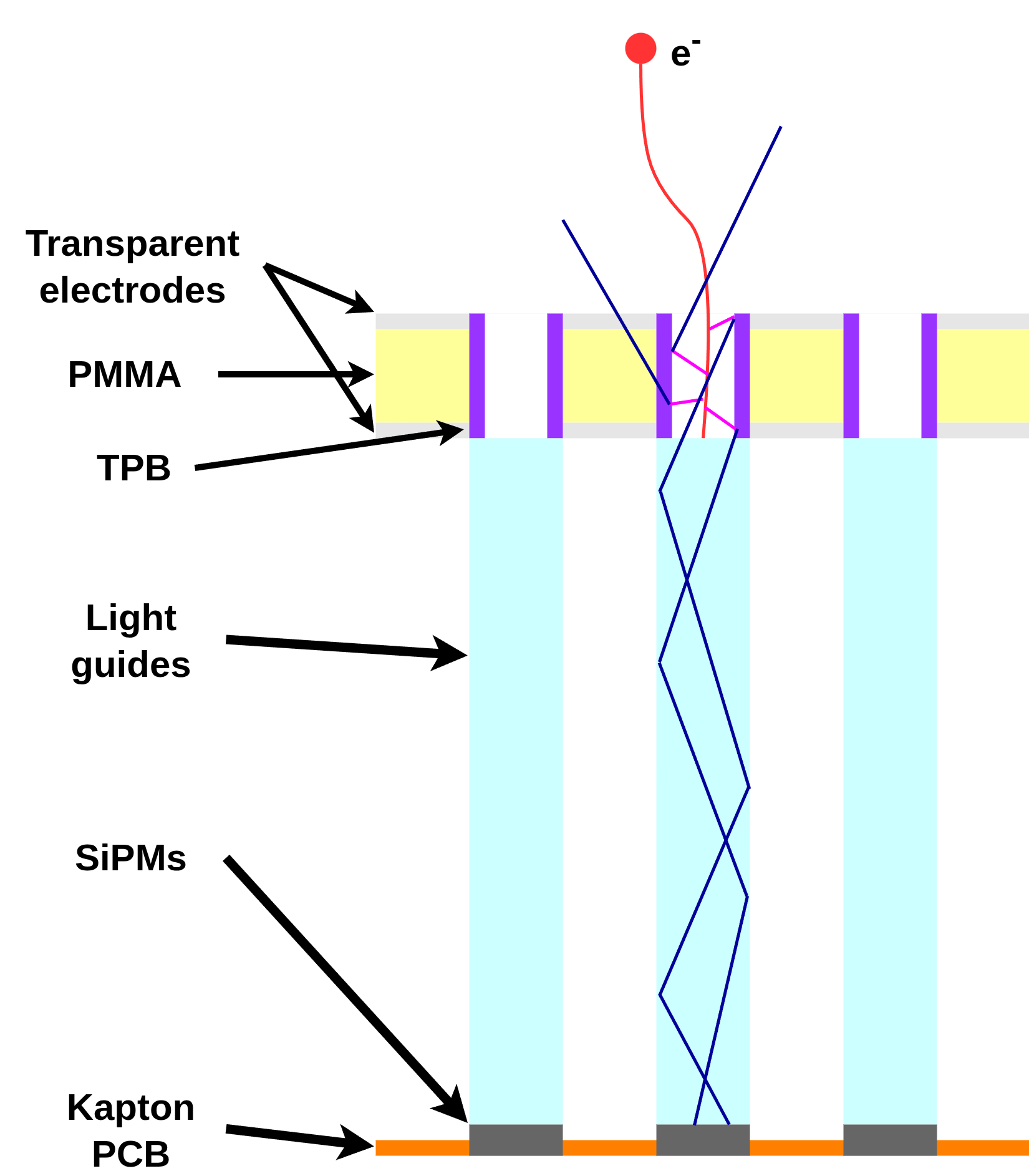}
    \caption{Principle of operation of the ELMAS. Drift electrons are driven into the FAT-GEM by the intense electric field (about \SI{2}{kV/cm/bar}) following the field lines shown in the top-right panel. Inside the channels, electrons produce VUV electroluminescence. The light is subsequently wavelength-shifted into the visible range through TPB-coating of the holes' walls. Part of the light going backwards is detected by the BFD. The light exiting the FAT-GEM channels is directed towards the Light Guide Honeycomb, which transports it to the DSP.}
    \label{fig:fatgem}
\end{figure}
 
 A large-diameter HPXeEL needs to find an alternative solution to 
  the EL grids used by current NEXT TPCs which are very hard to build beyond diameters of about \SI{1}{m}. For ITACA we propose the Electroluminescence Modular Amplification Structure (ELMAS), a set of tiles consisting of a Field Assisted Transparent Gas Electron Multiplier (FAT-GEM)~\cite{Leardini_2024} structures coupled to a Light Guide Honeycomb.

Figure~\ref{fig:fatgem} shows the principle of operation. The FAT-GEM is fabricated in a PMMA substrate \SI{5}{mm} thick. The PMMA is coated with resistive ITO and holes of \SI{6}{mm} diameter at a pitch of \SI{10}{mm} are laser-drilled into the substrate, defining the amplification channels. The walls of the holes are coated with TPB, so that VUV light impinging on them is wavelength-shifted to blue. A potential difference of \SI{16}{kV} is applied between the two electrodes, resulting, at \SI{15}{bar}, in a reduced field of \SI{2.1}{kV/cm/bar}, which produces approximately 1500~VUV photons per electron and approximately 1000~visible photons per electron. As shown in~\cite{Leardini_2024}, this structure has a performance approaching that of parallel grids, but permits the construction of a tiled EL region. Forward-moving photons are transported to the DSP (sitting at ground) by  
the Light Guide Honeycomb. The rods of the device are \SI{9.5}{cm} long, resulting in an average electric field of \SI{4.89}{kV/cm} across the structure. At \SI{15}{bar}, the corresponding reduced field 
is $4.89/15 \approx \SI{0.33}{kV\,cm^{-1}\,bar^{-1}}$, well below the EL production threshold ($\sim\SI{0.8}{kV\,cm^{-1}\,bar^{-1}}$) and far from breakdown, ensuring purely passive voltage degradation. The Light Guide retains approximately 45\% of the captured photons. Given the large coverage and small pitch between sensors in the DSP, this guarantees good tracking and provides a sub-percent measurement of the energy.

\subsubsection*{Dense Silicon Plane}


\begin{table}[htbp]
\centering
\begin{tabular}{llrl}
\hline
\textbf{Parameter} & \textbf{Symbol} & \textbf{Value} & \textbf{Unit} \\
\hline
\multicolumn{4}{l}{\textit{Input Parameters}} \\
DSP diameter & $D_\mathrm{DSP}$ & 320 & cm \\
SiPM pitch & $p$ & 10.0 & mm \\
SiPM size & --- & $6 \times 6$ & mm$^2$ \\
\hline
\multicolumn{4}{l}{\textit{Computed Values}} \\
Number of channels & $N_\mathrm{ch}$ & 80,381 & --- \\
Coverage & $f_\mathrm{cov}$ & 36.0 & \% \\
DSP total area & $A_\mathrm{DSP}$ & 8.04 & m$^2$ \\
SiPM active area & $A_\mathrm{SiPM}$ & 2.89 & m$^2$ \\
\hline
\multicolumn{4}{l}{\textit{System Estimates}} \\
Power per channel & $P_\mathrm{ch}$ & 5 & mW \\
Total power & $P_\mathrm{tot}$ & 402 & W \\
\hline
\end{tabular}
\caption{Dense Silicon Plane (DSP) parameters.}
\label{tab:dsp}
\end{table}

The Dense Silicon Plane (DSP) is mounted on a thin Kapton foil carrier board, similar to those used by NEXT. It operates at ground potential. A reasonable compromise between track separation and cost (number of channels) is achieved by pixelating the plane with silicon detectors of  $6 \times 6$~mm$^2$ at a pitch of \SI{10}{mm}. The SiPMs are read out by ASIC electronics, such as those being developed by the NEXT collaboration. Table \ref{tab:dsp} shows the main parameters. 

\subsubsection*{Barrel Fibre Detector}
 
\begin{figure}[!htbp]
    \centering
    \includegraphics[scale=0.30]{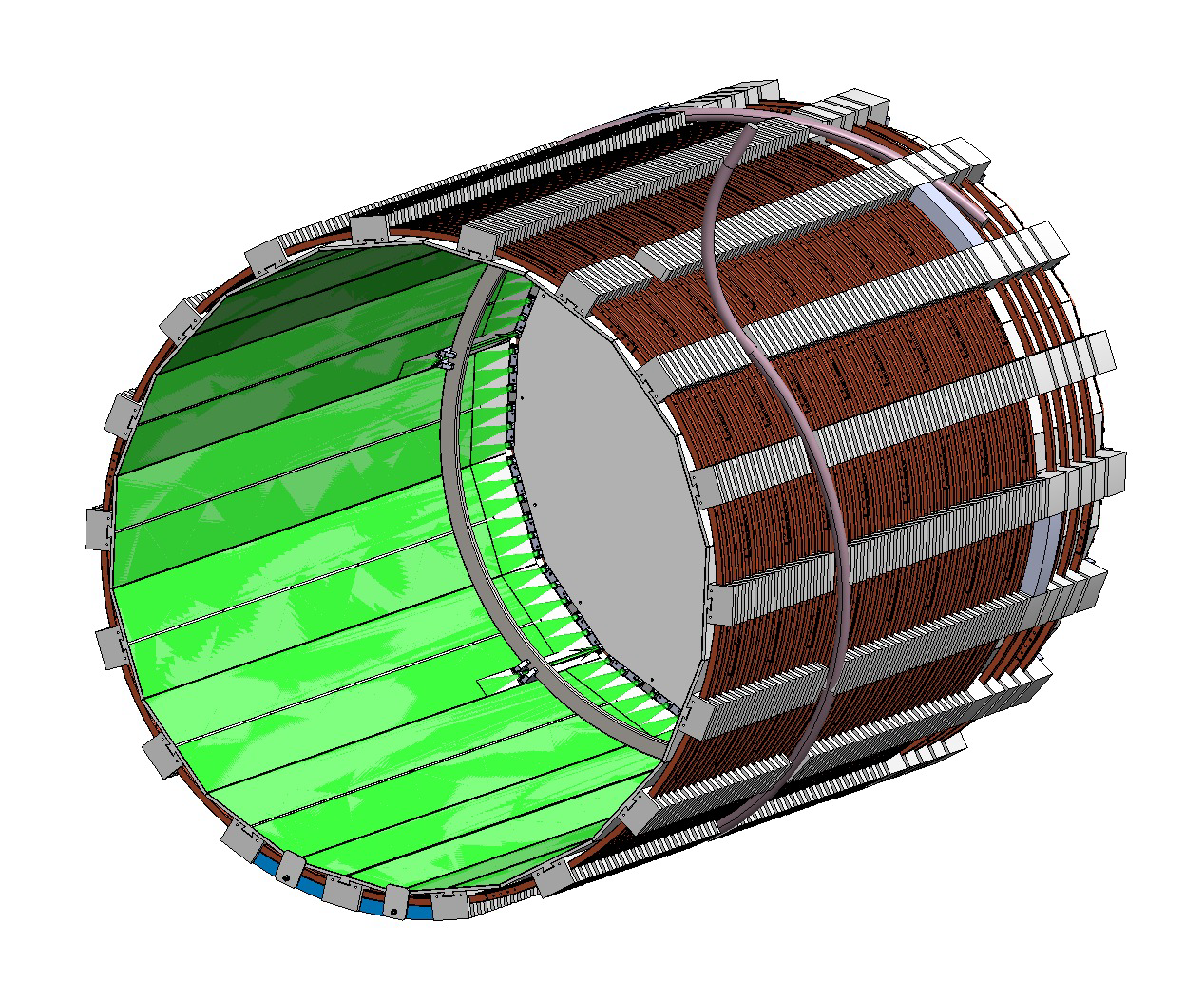}
    \caption{Conceptual design of the BFD~\cite{Soleti2024FiberBarrel}.}
    \label{fig:bfd}
\end{figure}

A conceptual drawing of the central detector, including field cage and the BFD, described in~\cite{Soleti2024FiberBarrel}, is shown in Figure~\ref{fig:bfd}. Optical fibres of \SI{1}{mm} diameter, assembled in PTFE panels and coated with TPB, are placed in a barrel around the field cage. The light from the fibres is read out by SiPMs, whose dark noise is suppressed by placing them in contact with a cooled copper ring. The BFD is able to measure primary scintillation \sone\ (and thus the start-of-the-event $t_0$), as well as \stwo. The moderate length of the TPC results in modest light attenuation in the fibres. 

\subsection{Principle of operation}
\label{sec:pof}

Consider a putative \bbonu\ event occurring in the middle of the TPC. Two electrons, with a total energy of \Qbb\ (\SI{2458}{keV}), propagate through the dense noble gas, ionising it. Primary scintillation is recorded by the BFD, and the Fast Trigger System (FTS) records the start-of-the-event $t_0$.

The ionisation electrons reach the anode in about \SI{0.7}{ms}. The FTS then performs a fast measurement of the event energy, accepting the event only if its energy lies within a pre-determined ROI (selected to be approximately 1~FWHM around \Qbb). Events outside the ROI are rejected and no ion readout is attempted.

Since the event that we are considering is a \bbonu\ decay, the FTS will likely trigger, activating a fast reconstruction of the topology and event barycenter, which in turn permits the prediction of the barycenter of the ion track at the cathode, $(x_i, y_i, z_i)$. Using $t_0$, the FTS also predicts the time of arrival of the ion cloud ($t_i$). All these operations can be performed in less than \SI{500}{ms}. 
     
The \xetp\ ions reach the cathode on a time scale of seconds. In our specific example, with the event at the midpoint of the TPC, they need to drift \SI{75}{cm} at a velocity of \SI{10}{cm/s}, requiring \SI{7.5}{s}. The long drift time permits displacing the ion detector to $(x_i, y_i, z_i)$ and opening an electrostatic gate in the ion detector from $t_i$ to $t_i + \Delta t$, where $\Delta t$ is the time needed to collect the full ion cloud. The typical longitudinal extension of the ion track is of the order of \SI{10}{cm}, and thus about \SI{1}{s} is needed to collect the ions. See section~\ref{sec:mars} for further discussion.

Notice that the electron track and the ion track are detected in \emph{delayed coincidence}. The ion track is intercepted precisely at the right time and in the right place, thus minimising the impact of spurious coincidences. At the expected trigger rate (dominated by backgrounds in the ROI), pile-up during the $\sim$\SI{15}{s} maximum ion drift time is negligible.

\section{The MARS System}
\label{sec:mars}

\subsection{Concept}
In the ITACA detector the ion track is captured by a small CMOS sensor that must be positioned at the predicted arrival point $(r_0,\theta_0)$ before the ions reach the cathode. This is achieved with the Magnetically Actuated Rotor System (MARS).

\begin{figure}[!htbp]
    \centering
    \includegraphics[scale=0.5]{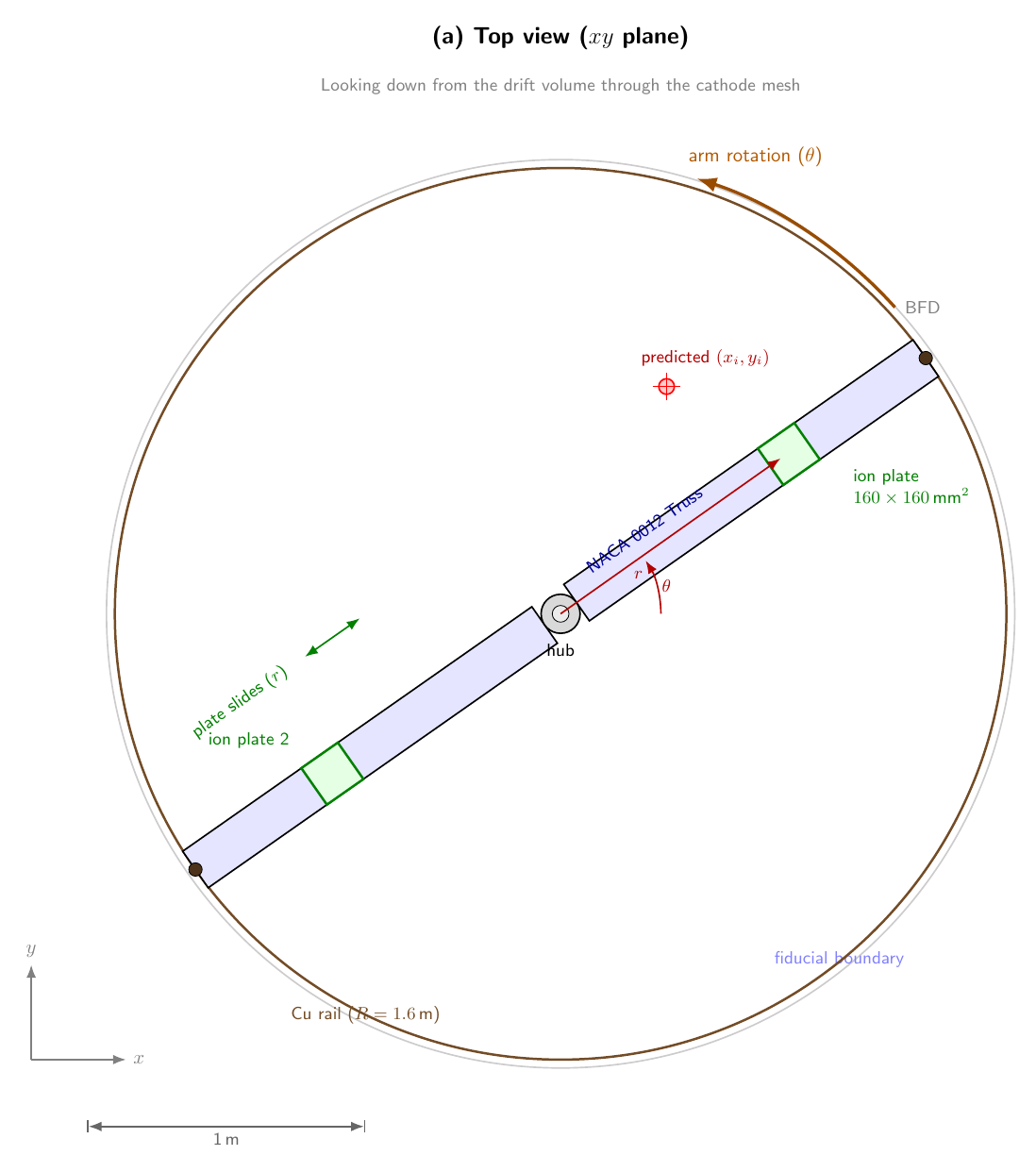}
    \caption{XY view of the MARS system, showing the dual-arm propeller, copper rail, and ion plate positions.}
    \label{fig:marsXY}
\end{figure}

\begin{figure}[!htbp]
    \centering
    \includegraphics[scale=0.3]{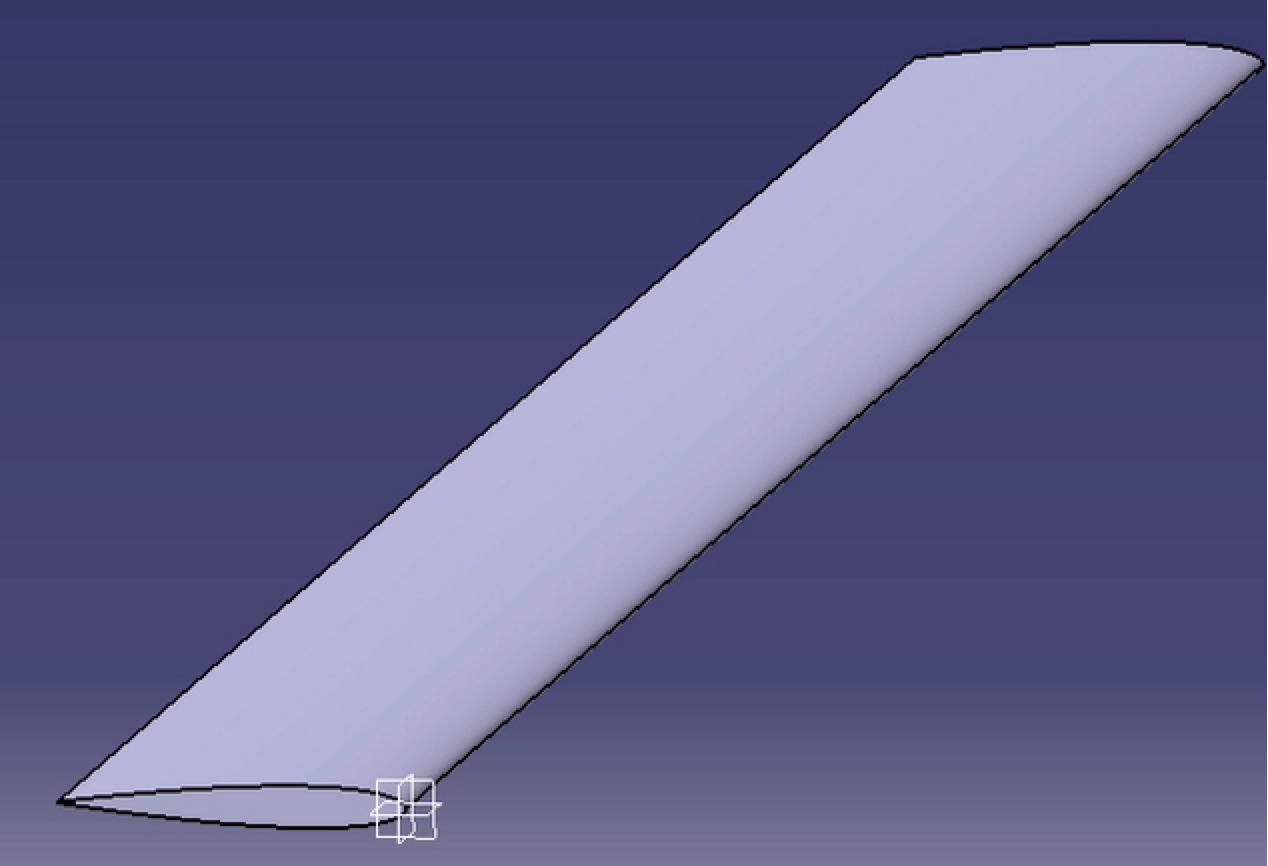}
    \caption{A 3D rendering of the propeller arm with the NACA\,0012 profile.}
    \label{fig:naca3d}
\end{figure}

MARS positions the ion detector at any coordinate $(r,\theta)$, a few mm below the cathode plane. This requires: 1) an azimuthal sweep, performed by two arms which rotate from a central hub and slide through rails that run around the detector diameter; 2) a radial sweep which is performed by the ion detector carrier. Figure~\ref{fig:marsXY} shows a diagram of the XY projection of the system.

The challenge here is to perform the sweep rapidly through dense xenon gas (to reduce dead time), while minimising gas disturbance that could deflect the slowly drifting ions from their intended collection points.  As we will discuss below, the low viscosity of xenon plays a major role in the formation and propagation of those gas disturbances, which, however, are effectively damped by the ion focusing grid (IFG), a structure that plays a dual role: a) it blocks the gas disturbances introduced by the blade movement; and b) it focuses the  incoming ions into the reading pads of the CMOS detector.

Mechanically, the system must satisfy four strict constraints: 1) there can be no oils or lubricants inside the chamber; 2) moving parts are actuated magnetically to avoid feedthroughs; 3) all the sliding parts must minimise friction and 4) all components must be radiopure.
\subsection{Geometry}

\begin{table}[ht]
\centering

\begin{tabular}{lrl}
\toprule
Parameter & Value & Unit \\
\midrule
Arm half-span $R$ & 1.6 & \si{\metre} \\
Arm chord $c$ (tangential) & 160 & \si{\milli\metre} \\
Profile & \multicolumn{2}{l}{NACA\,0012 (symmetric top/bottom)} \\
Max.\ thickness $t_\text{max}$ & 19.2 & \si{\milli\metre} \\
Linear mass density $\mu_\text{arm}$ & 0.28 & \si{\kilo\gram\per\metre} \\
Ion-plate mass $m_\text{plate}$ & 0.25 & \si{\kilo\gram} \\
Motor torque $\tau_\text{motor}$ & 60 & \si{\newton\metre} \\
Number of arms & 2 & --- \\
Rotation direction & \multicolumn{2}{l}{Unidirectional (LE first)} \\
Maximum rotation $\Delta\theta$ & $\pi$ & \si{\radian} \\
\bottomrule
\end{tabular}
\caption{Parameters of the NACA blade.}
\label{tab:mars_params}
\end{table}

MARS employs a propeller geometry, e.g., two helicopter-like blades at zero angle of attack. Each arm is a hollow NACA\,0012 profile, shown in Figure~\ref{fig:naca3d}.
The cross-section of the profile lies in the vertical plane: the chord ($c = 160$~mm) is horizontal along the tangential (sweep) direction, the maximum thickness ($t_\text{max} = 0.12\,c = 19.2$~mm) is vertical, and the span ($R = 1.6$~m) is radial.  The chord is chosen to match the tangential dimension of the $160\times 160$~mm$^2$ ion detector, so that the plate is fully contained within the profile. Ions drift vertically downward onto the broad, flat top surface of the arm. Because the NACA\,0012 has a rounded leading edge and a sharp trailing edge, the arm must always rotate leading-edge first for optimal aerodynamic performance.  With two arms at $180^\circ$ spacing, each arm covers a semicircle, and the worst-case rotation to reach any point is $\Delta\theta = \pi$.

The symmetric NACA profile at zero angle
of attack generates no lift and minimises drag. The hollow interior
accommodates the ion detector, ion carrier mechanism and cabling. The blade skeleton is made of three titanium tubes arranged in a triangular configuration at vertical positions
within the NACA\,0012 profile. A number of HDPE ribs define the airfoil shape and support the Kapton skin. The blade upper (and lower) surfaces are designed as a conveyor belt, attached at its ends to the ion carrier detector. As the carrier slides along the radial coordinate, the belt tracks with it, maintaining an unbroken aerodynamic surface.

The ion detector (ID) consists of two components, the ion detector carrier (IDC), and the ion collection plate (ICP). The IDC slides radially within the blade, driven by a spindle pushing the Kapton conveyor belt. The ICP contains the CMOS sensors. Once the IDC is positioned in the target position, the ICP lifts vertically to a docking position below the IFG.

\subsection{Xenon gas at \SI{15}{bar}}

\begin{table}[htbp!]
\centering
\caption{Properties of xenon gas at \SI{15}{bar}, \SI{300}{K}.}
\label{tab:xe_props}
\begin{tabular}{lrl}
\toprule
Property & Value & Unit \\
\midrule
Pressure $P$ & 15 & bar \\
Temperature $T$ & 300 & \si{\kelvin} \\
Density $\rho$ & 85.9 & \si{\kilo\gram\per\cubic\metre} \\
Dynamic viscosity $\mu$ & $2.32\times 10^{-5}$ & \si{\pascal\second} \\
Kinematic viscosity $\nu$ & $2.70\times 10^{-7}$ & \si{\metre\squared\per\second} \\
Speed of sound $c_s$ & 178 & \si{\metre\per\second} \\
\bottomrule
\end{tabular}
\end{table}

At $P = 15$~bar and $T = 300$~K, xenon has the properties listed in Table~\ref{tab:xe_props}.  The kinematic viscosity $\nu = \mu/\rho \approx 2.70\times 10^{-7}$~\si{m^2/s} is roughly 56 times smaller than that of air at atmospheric pressure ($\nu_\text{air} \approx 1.5\times 10^{-5}$~\si{m^2/s}). This has two consequences. First, it induces a turbulent boundary layer (BL) around the blade when it moves. Second, the diffusion of this BL is extremely slow.

The speed of sound in xenon at these conditions is $c_s \approx 178$~\si{m/s}, so acoustic adjustment over the detector diameter takes ${\sim}\,$18~ms---effectively instantaneous on the timescales considered here.

At a drift field of \SI{200}{V/cm}, $\text{Xe}_2^+$ ions drift at $v_\text{ion} = 100$~\si{mm/s}~\cite{Gomez-Cadenas:2025ect}.  The maximum drift time over the \SI{1.5}{m} drift length is $t_\text{drift} = 15$~s.

\subsection{Motion Sequence}

The positioning consists of three sequential phases, each driven
by an independent actuator. Sequential execution simplifies the mechanical
design by decoupling the drive shafts and allows clean kinematics for each
phase.

\subsubsection*{Phase 1: Blade Rotation}

\begin{table}[bthp!]
\centering
\caption{Blade Rotation.}
\label{tab:kin-rot}
\begin{tabular}{lll}
\toprule
Quantity & Value & Notes \\
\midrule
Sweep angle       & $\theta = 180^\circ$              & $= 2\pi/2$ \\
Rotation time     & $t_\mathrm{rot} = 395$~ms          & Bang-bang \\
Peak angular velocity & $\omega_\mathrm{max} = 15.9$~rad/s & At midpoint \\
Tip velocity      & $V_\mathrm{tip} = 25.4$~m/s         & $= \omega_\mathrm{max} R$ \\
Mach number       & $\mathrm{Ma} = 0.14$                & Incompressible \\
\bottomrule
\end{tabular}
\end{table}

The arms rotate through a maximum sweep angle of $\pi$. The total moment of inertia includes blades and carrier plates, which are accounted for as point masses at radius $r_0$:
\begin{equation}
  I_\mathrm{total} = n \left(
    \frac{1}{3}\,m_\mathrm{blade}\,R^2 + m_\mathrm{plate}\,r_0^2
  \right)
  \label{eq:I-total}
\end{equation}
with $m_\mathrm{blade} = 0.25$~kg, $m_\mathrm{plate} = 0.25$~kg,
$R = 1.6$~m, and $r_0 = 0.8$~m.

For motor torque $\tau = 60$~N$\cdot$m and bang-bang kinematics:
\begin{equation}
  t_\mathrm{rot} = 2\sqrt{\frac{\theta\,I}{\tau}}, \qquad
  \omega_\mathrm{max} = \sqrt{\frac{\theta\,\tau}{I}}, \qquad
  V_\mathrm{tip} = \omega_\mathrm{max}\,R
  \label{eq:bangbang-rot}
\end{equation}

\subsubsection*{Phase 2: Carrier Slide}
\begin{table}[bthp!]
\centering
\caption{Carrier slide.}
\label{tab:kin-slide}
\begin{tabular}{lll}
\toprule
Quantity & Value & Notes \\
\midrule
Slide distance    & $\Delta r = 0.8$~m              & \\
Slide time        & $t_\mathrm{slide} = 200$~ms     & Bang-bang \\
Peak slide velocity & $V_\mathrm{slide} = 8$~m/s  & Internal to blade \\
\bottomrule
\end{tabular}
\end{table}
After rotation, the carrier slides radially by a maximum value of $\Delta r = 0.8$~m inside
the hollow blade. The external surface (Kapton belt + plate) remains
stationary relative to the gas. For actuator force $F = 20$~N
and carrier mass $m = 0.25$~kg, bang-bang kinematics gives:
\begin{equation}
  t_\mathrm{slide} = 2\sqrt{\frac{\Delta r\,m}{F}}, \qquad
  V_\mathrm{slide} = \sqrt{\frac{\Delta r\,F}{m}}
  \label{eq:bangbang-slide}
\end{equation}

\subsubsection*{Phase 3: Vertical Lift}

After the slide, the ICP carrying the CMOS detectors lifts vertically by
$z_\mathrm{lift} = 10$~mm over $t_\mathrm{lift} = 0.5$~s, until it reaches its docking position below the IFG.
\subsubsection*{Total Cycle Time}

The three phases execute sequentially:
\begin{equation}
  t_\mathrm{total} = t_\mathrm{rot} + t_\mathrm{slide} + t_\mathrm{lift}
    = 395 + 200 + 500~\text{ms}
    \approx \SI{1.1}{s}
  \label{eq:t-total}
\end{equation}

Ions produced within a distance $d_\mathrm{dead} = v_d \times t_\mathrm{total}$
of the cathode arrive at the collection plane before the positioning sequence
is complete, and are therefore lost. This \emph{dead zone} amounts to:
\begin{equation}
  d_\mathrm{dead} = \SI{100}{mm/s} \times \SI{1.1}{s}
    = \SI{110}{mm}
    \approx 7.3\%~\text{of the drift length}
\end{equation}


\subsection{Gas Disturbance Mechanisms}
\label{sec:mechanisms}

Moving mechanical components through dense xenon induces flow disturbances.
Below we discuss the dominant mechanisms by which MARS motion could
affect ion trajectories.

\subsubsection*{Blade Boundary Layer}

The rotating blade drags gas along its upper surface, forming a turbulent
boundary layer. The chord Reynolds number is
\begin{equation}
  \mathrm{Re}_c = \frac{V_\mathrm{tip}\,c}{\nu} = 1.51\times 10^{7}
\end{equation}
with chord $c = 160$~mm. Since $\mathrm{Re}_c \gg 5 \times 10^5$,
the boundary layer is fully turbulent.

\paragraph{Boundary layer thickness.}
The Prandtl--Schlichting flat-plate correlation gives:
\begin{equation}
  \boxed{\delta = 0.37\,L\,\mathrm{Re}_L^{-1/5}}
  \label{eq:bl-thickness}
\end{equation}
Applied with $L = c$ and $V = V_\mathrm{tip}$, we obtain that the initial thickness of the BL associated with the blade is:
$\delta_\mathrm{blade} = 2.17$~mm.

The flow is purely tangential, with a mean velocity:
\begin{equation}
  u_\theta(z) = V_\mathrm{tip}
    \left(\frac{z}{\delta_\mathrm{blade}}\right)^{1/7}
    \quad \text{for } z < \delta_\mathrm{blade}
  \label{eq:u-theta}
\end{equation}

\paragraph{Turbulent transport after motion stops.}
When the blade stops, the mean flow within $\delta$ persists, and the turbulent eddies that were superimposed on the mean flow continue transporting momentum vertically for a brief period
before decaying. The outer-layer eddies --- the largest structures, and the
ones responsible for vertical transport beyond $\delta$ --- have a
characteristic fluctuation velocity $u' \approx 0.05\,V_\mathrm{tip}
= \SI{1.27}{m/s}$ (a standard empirical ratio for turbulent boundary
layers) and decay with a timescale of
$t_\mathrm{decay} \approx \delta / u' \approx 1.7$~ms. The turbulent transport distance is
$\ell_\mathrm{turb} = u' \times t_\mathrm{decay}$. Thus,
the maximum vertical extent of gas disturbance shortly after the blade stops is the boundary layer thickness plus the distance eddies can transport momentum before decaying:
\begin{equation}
  \boxed{z_\mathrm{max} = \delta + \ell_\mathrm{turb}
    = \SI{4.34}{mm}}
  \label{eq:max-reach-blade}
\end{equation}

\subsubsection*{Bulk Swirl}
\label{sec:m3}

During the sweep the blade surface moves tangentially through the gas at up to
$V_\mathrm{tip} = 25.4$~m/s. Viscous friction at the surface drags the
adjacent gas along, spinning it up: each fluid parcel within the boundary
layer acquires tangential velocity and therefore angular momentum about the
central axis. By the time the blade stops, a layer of gas of thickness
$\delta_\mathrm{blade} = 2.17$~mm is co-rotating with the blade. This rotating gas layer constitutes the bulk swirl.

At the moment the blade stops, the swirl layer has thickness
$\delta_{\mathrm{swirl},0} = \delta_\mathrm{blade} = 2.17$~mm.

\subsubsection*{Carrier Boundary Layer}
\label{sec:m2}

In a design where the carrier slides externally, it would create a radial
boundary layer analogous to the blade BL but with $V = V_\mathrm{slide} = 8$~m/s
and characteristic length $L = c$. In MARS, however,
the carrier slides inside the hollow blade, and a flat surface is always presented to the gas thanks to the conveyor belt. Thus, the associated (radial) BL is suppressed with respect to the BL created by the blade rotation.

\subsubsection*{Stokes Layer from Vertical Lift}
\label{sec:stokes}

After the carrier reaches the target point, the ICP carrying the CMOS sensors is lifted slowly, until it ``docks'' with the IFG.
This creates a vertical velocity field.
\begin{equation}
  \mathrm{Re}_\mathrm{lift}
    = \frac{V_\mathrm{lift}\,z_\mathrm{lift}}{\nu}
    = 1481 \quad (\text{laminar})
\end{equation}
The viscous penetration depth is a thin Stokes layer:
\begin{equation}
  \delta_\mathrm{Stokes} = \sqrt{\nu\,t_\mathrm{lift}}
    = \SI{0.37}{mm}
  \label{eq:stokes}
\end{equation}

This is a purely \emph{vertical} velocity --- it only has a tiny effect on ion arrival timing
of the order of $\Delta t \sim \delta_\mathrm{Stokes}/v_d \sim 4$~\si{\micro\second}, which is
negligible on the 1--\SI{15}{s} drift timescale.

\subsubsection*{Transient effects}

\paragraph{Potential flow:} the moving blades displace gas, creating an irrotational (potential) flow
field that extends throughout the chamber. This flow adjusts on the acoustic
timescale:
\begin{equation}
  t_\mathrm{acoustic} = \frac{R}{c_s} = \SI{9}{ms}
  \label{eq:t-acoustic}
\end{equation}

When the blade stops, the potential flow vanishes
within a few acoustic times:
\begin{equation}
  t_\mathrm{decay} \approx 2\,t_\mathrm{acoustic} = \SI{18}{ms}
\end{equation}
%
\paragraph{Pressure Pulse:} rapid blade acceleration generates a pressure pulse that propagates as an acoustic wave.
The pulse traverses the chamber in:
\begin{equation}
  t_\mathrm{traverse} = \frac{2\,R_\mathrm{cyl}}{c_s}
    = \SI{18}{ms}
  \label{eq:t-traverse}
\end{equation}
and decays within $\sim 3\,t_\mathrm{traverse} \approx 54$~ms via
reflections and viscous absorption.

\subsection{Persistent gas disturbance}

In summary, the positioning of the ion detector creates two persistent mechanisms which produce a horizontal flow: the BL associated to the blade sweep and the bulk swirl. The maximum vertical reach of the disturbed gas in both cases (at $t = 0$) is of the order of \SI{4.3}{mm}.
Both produce purely horizontal (tangential) velocity of large
magnitude ($V_\mathrm{tip} = 25.4$~m/s at the blade
surface). To avoid that ions traverse
this turbulent/swirling region, the CMOS detector is lifted by the ICP. The minimum lift needed is
$z_\mathrm{lift}^\mathrm{min} \sim 4.3$~mm, to fully avoid the turbulent region. Notice, however, that ions traverse the chamber in a time scale of many seconds, with a maximum drift time of
\SI{15}{s}. During this time, the BL layers associated to the blade and the swirl will diffuse horizontally, potentially reaching $z_\mathrm{lift}^\mathrm{min}$, which needs to be adjusted accordingly.

Importantly, the BL layers continue their slow diffusion for a very long time, potentially affecting other events that may occur \emph{after} the event that has triggered the sweep. To understand both these short and long term effects we need to analyse the effects of diffusion of the BL layer(s).

Dedicated simulations performed for the NEXT-100 detector~\cite{NEXT:2025yqw} show that the gas recirculation induces velocities $<1$~mm/s, predominantly along the $z$ direction. Given that this is two orders of magnitude smaller than the ion drift velocity ($\sim100$~mm/s), and largely parallel to it, the resulting effect on ion trajectories, particularly in the transverse plane, is negligible. We therefore conclude that recirculation-driven flows do not constitute a concern for the ion imaging performance.

\subsection{Diffusion of the BL layers}
\label{sec:diff-model}

The initial boundary layer profiles for both the sweep and the swirl (Eq.~\ref{eq:u-theta})
can be approximated as a momentum-conserving slab of uniform velocity
$\sim V_\mathrm{tip}$ and width $\delta$.

After the blade stops, the surface is stationary
and the velocity evolves by one-dimensional viscous diffusion:
\begin{equation}
  \boxed{u(z,t) = \frac{V_\mathrm{eff}}{2}
    \left[2\,\mathrm{erf}\!\left(\frac{z}{2\sqrt{\nu t}}\right)
          - \mathrm{erf}\!\left(\frac{z - \delta}{2\sqrt{\nu t}}\right)
          - \mathrm{erf}\!\left(\frac{z + \delta}{2\sqrt{\nu t}}\right)
    \right]}
  \label{eq:u-diffusion}
\end{equation}

\begin{figure}[!htbp]
    \centering
    \includegraphics[width=\textwidth]{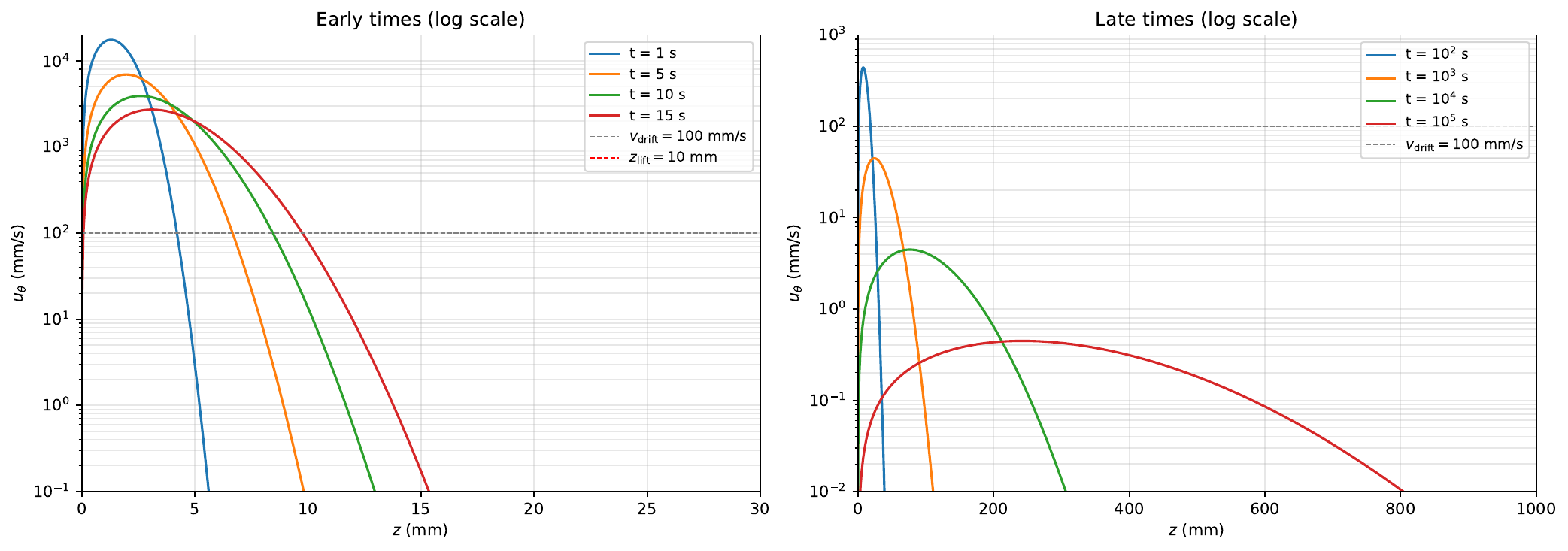}
    \caption{Velocity profiles $u_\theta(z)$ at selected times after motion stops.
    Left: early times (\SI{1}{s} to \SI{15}{s}), log scale.
    Right: late times ($10^2$ to $10^5$~s), log scale in \si{mm/s}.
    The dashed line marks $v_\mathrm{drift} = 100$~mm/s;
    the vertical red line marks $z_\mathrm{lift} = 10$~mm.}
    \label{fig:diffbl}
\end{figure}

The equation describes both the spreading upwards of the BL layer, as
momentum diffuses into the
quiescent region ---with the BL reach growing as
$z_\mathrm{BL} \sim \delta + \sqrt{\nu\,t}$ --- and the damping of the velocity as
the fixed initial momentum is
redistributed over a growing layer.

Figure~\ref{fig:diffbl} shows the distribution of the $u_\theta$ velocity as a function of $z$. The left panel considers what happens for short times ($t \leq 100$~s).
Notice that for $t = 15$~s, the profile reaches \SI{10}{mm}, but the lateral velocities are of the same order as the drift velocity while the crossing time is very small, resulting in lateral displacement well below the effect of the diffusion. This is the reason for lifting the ICP by $z_\mathrm{lift} = 10$~mm, ensuring that even the ions that are produced very near the anode reach the CMOS without being substantially deflected.

The right panel shows $u_\theta$ for long times ($t \leq 10^5$~s). For relatively short times ($t \sim 100$~s), $u_\theta$ peaks above the ion drift velocity, while for $t \geq 10^3$~s it is always below $v_\mathrm{drift}$.

\begin{figure}[!htbp]
    \centering
    \includegraphics[width=\textwidth]{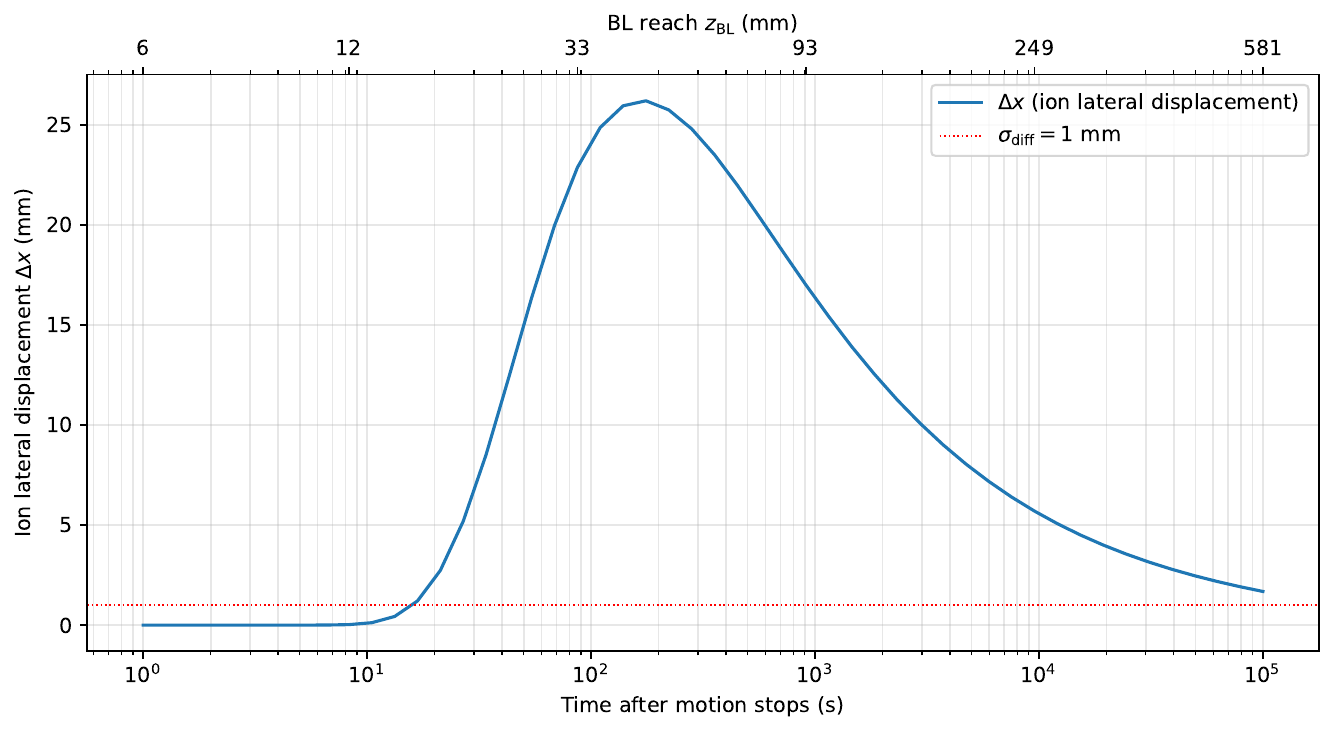}
    \caption{Ion lateral displacement as a function of time after motion stops.
    The red dotted line marks $\sigma_\mathrm{diff} = 1$~mm.}
    \label{fig:idx}
\end{figure}

The velocity field introduces a displacement in ions propagating at later times after the event that has triggered the MARS motion. Figure~\ref{fig:idx} shows this displacement as
a function of time after motion stops. Notice that the displacement is small for $t \leq 15$~s ---that is, times within the same event---, then peaks at about \SI{200}{s}, with a large displacement of $\sim\!\SI{25}{mm}$, and does not reach a value of the same order as the diffusion until about $10^5$~s.

In principle, this lateral displacement could be corrected, but it would require precise knowledge of the velocity fields and/or a careful calibration. A more robust solution is available, as discussed below.

\subsection{The Ion Focusing Grid}
\label{sec:ifg}

The Ion Focusing Grid (IFG) is a modular structure similar to the Electroluminescence Modular Amplification Structure deployed in the anode. It consists of a tightly packed set of FAT-GEM tiles fully covering the cathode plane. Like its cousin in the anode, the FAT-GEM is fabricated in a PMMA substrate coated with resistive ITO or with a conductive copper layer. The dimensions, however, are different, to adapt to its dual role as gas disturbance barrier and ion focusing grid. In this case the thickness is \SI{10}{mm}, the holes have a diameter of \SI{0.5}{mm} and the pitch is \SI{2}{mm}.

The IFG is located in the cathode plane.
The open area fraction is:
\begin{equation}
  f = \frac{\pi\,(d_h/2)^2}{p^2}
    = \frac{\pi \times 0.25^2}{4} \approx 0.049
    \quad (4.9\%)
  \label{eq:icg-f}
\end{equation}
and the aspect ratio:
\begin{equation}
  \frac{L_\mathrm{IFG}}{d_h} = \frac{10.0}{0.5} = 20
  \label{eq:icg-aspect}
\end{equation}

\subsubsection*{The Ion Focusing Grid as a Gas Disturbance Barrier}

The diffusing boundary layer is horizontal momentum propagating vertically
by molecular viscosity. When this momentum reaches the IFG, two effects
combine to block its transmission: 1) the ITO/metal surface of the IFG is a solid
barrier that absorbs incoming horizontal
momentum. This accounts for 95\% of the
grid area; 2) in the remaining 5\%, horizontal momentum enters the holes. Each hole is a
cylindrical channel of length $L_\mathrm{IFG} = 10$~mm and diameter \SI{0.5}{mm}, with
no-slip walls. Horizontal velocity entering the channel decays due to
viscous friction on the walls. For a velocity profile diffusing through a
narrow channel, the attenuation scales as:
\begin{equation}
  \frac{u_\mathrm{exit}}{u_\mathrm{entry}}
    \sim \exp\!\left(-\frac{L_\mathrm{IFG}}{d_h}\right)
    = e^{-20} \approx 2 \times 10^{-9}
  \label{eq:channel-atten}
\end{equation}
Therefore, the gas disturbance is totally blocked by the IFG.

\subsubsection*{Ion Focusing onto Detector Pads}
\label{sec:icg-focusing}

Beyond blocking gas disturbance, the IFG serves a second, essential function: it
focuses drifting ions through its holes and onto the CMOS detector pads. As we will discuss in section \ref{sec:nausica}, this focusing is essential to detect the ions properly.

To achieve the focusing, the top electrode of the IFG (the gate) is set at a
voltage of $V_\mathrm{gate} = 1200$~V, and the bottom electrode (the ion cathode or IC) at a voltage of
$V_\mathrm{IC} = 1100$~V, while the plate carrying the CMOS is
at ground ($V = 0$), and sits at a
distance $d_\mathrm{IC} = 5$~mm below the IFG.
The pads have the same diameter ($d_h = 0.5$~mm) and pitch
($p = 2$~mm) as the IFG holes.

\paragraph{Alignment:} When the plate lifts an integrated optical system allows the precise alignment of the CMOS pads with the IFG holes. The alignment system uses photodiodes embedded in the ion
  carrier to measure its position relative to the IFG hole pattern by detecting
  the contrast between reflection from the solid sheet in the bottom of the IFG (high signal) and transmission through holes (no signal). As the carrier displaces, each sensor
  sees a periodic reflectivity signal with the \SI{2}{mm} IFG pitch; the phase of this
  signal encodes the local offset with less than \SI{10}{\micro\metre} resolution (since 1\% intensity discrimination over a \SI{0.8}{mm} transition zone can be easily achieved). With 4 or more sensors at known
  positions across the \SI{160}{mm} carrier, the three rigid-body degrees of freedom
  are overdetermined with excellent resolution. The working principle is that the periodic hole
  pattern itself serves as the reference grating, a technique well developed for commercial
  optical encoders.

\paragraph{Electric fields:}
the drift field between the gate and the IC transports ions downward:
\begin{equation}
  E_\mathrm{gate} = \frac{V_\mathrm{gate} - V_\mathrm{IC}}{d_\mathrm{gate}}
    = \frac{\SI{100}{V}}{\SI{10}{mm}} = \SI{100}{V/cm}
  \label{eq:E-gate}
\end{equation}
While the collection field between the IC and the pads (ground) extracts ions through
the holes:
\begin{equation}
  E_\mathrm{IC} = \frac{V_\mathrm{IC}}{d_\mathrm{IC}}
    = \frac{\SI{1100}{V}}{\SI{5}{mm}} = \SI{2200}{V/cm}
  \label{eq:E-IC}
\end{equation}

\paragraph{Collection condition.}
Each unit cell of area $p^2$ funnels ions into a hole of area
$A_h = \pi(d_h/2)^2$. By flux conservation (Gauss's law), all field lines
entering a cell from above must pass through the hole. The required field
at the hole is:
\begin{equation}
  E_\mathrm{coll} = \frac{E_\mathrm{gate}}{f}
    = E_\mathrm{gate} \times \frac{p^2}{\pi\,(d_h/2)^2}
    = \SI{2037}{V/cm}
  \label{eq:E-coll}
\end{equation}
The collection condition is:
\begin{equation}
  \boxed{\frac{E_\mathrm{IC}}{E_\mathrm{gate}} \geq \frac{1}{f}
    \approx 20.4}
  \label{eq:coll-condition}
\end{equation}

The collection field of \SI{2200}{V/cm} gives a field ratio
$E_\mathrm{IC}/E_\mathrm{gate} = 22$, which exceeds the required ratio of 20.4.
Thus, ion field lines converge into the IFG holes and
every ion drifting downward through the gate is
funnelled onto the corresponding detector pad.


\subsection{Actuators}

\begin{figure}[!htbp]
    \centering
    \includegraphics[scale=0.8]{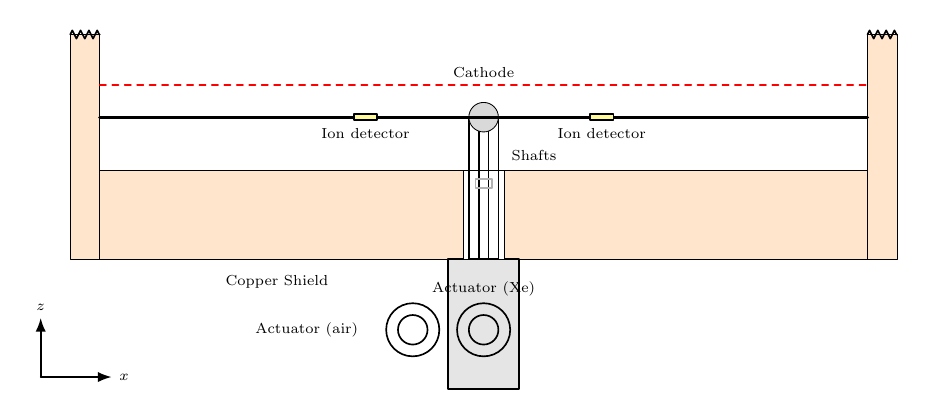}
    \caption{Schematics (not to scale) of the driving mechanism in MARS.}
    \label{fig:marsXZ}
\end{figure}

In MARS all actuation originates from below the copper floor shield,
preserving the radiopurity of the active volume. To avoid dynamic seals,
the system employs a dual-concentric, magnetic-activated design,
sketched in Figure~\ref{fig:marsXZ}:

\begin{itemize}
    \item \textbf{Re-entrant well:} A titanium well protrudes below the bottom flange, eliminating any rotating seal or feedthrough in the pressure boundary.
    \item \textbf{External stator (air side):} Two independent servo
          motors drive two concentric magnetic rings positioned around
          the outside of the titanium well.
    \item \textbf{Internal rotor (xenon side):} Inside the well, two
          corresponding magnetic rotors are nested concentrically. Each
          is magnetically locked to its external ring, allowing
          independent torque transmission through the static wall.
\end{itemize}

The drive-to-arm connection uses a nested shaft design that decouples
azimuthal rotation ($\theta$) from radial translation ($r$):

\begin{enumerate}
    \item \textbf{Outer shaft ($\theta$):} The outer magnetic rotor
          drives a hollow Ti shaft rising
          through a bore in the copper floor shield. It terminates in a
          central yoke to which the NACA\,0012 arms are bolted. Rotation
          of this shaft sets the $\theta$ coordinate.
    \item \textbf{Inner shaft ($r$):} The inner magnetic rotor drives a
          solid Ti shaft running concentrically
          inside the outer shaft. Above the yoke, a
          bevel gear redirects its rotation by $90^\circ$ into a spindle mechanism that
          moves the IDC.
\end{enumerate}


%

\section{NAUSICA for ITACA}
\label{sec:nausica}
\begin{figure}[!htbp]
    \centering
    \includegraphics[scale=0.40]{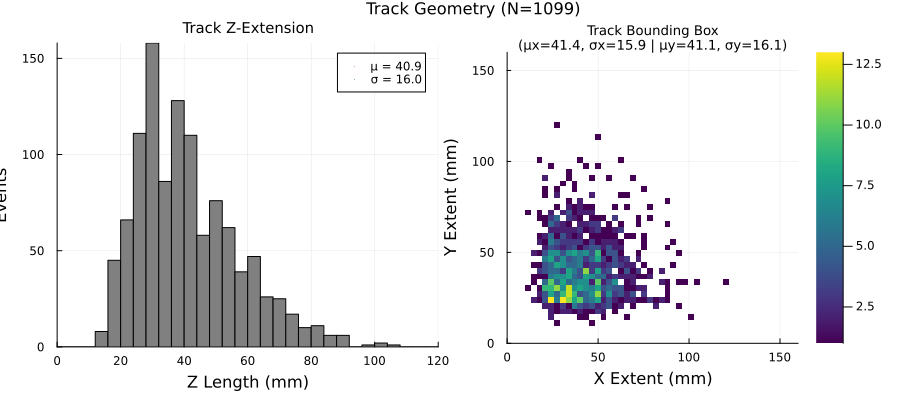}
    \caption{Left panel: z extension of \bbonu\ events; right panel: bounding box (maximum extension in x-y) of \bbonu\ events }
    \label{fig:zoc}
\end{figure}

\begin{figure}[!htbp]
    \centering
    \includegraphics[scale=0.40]{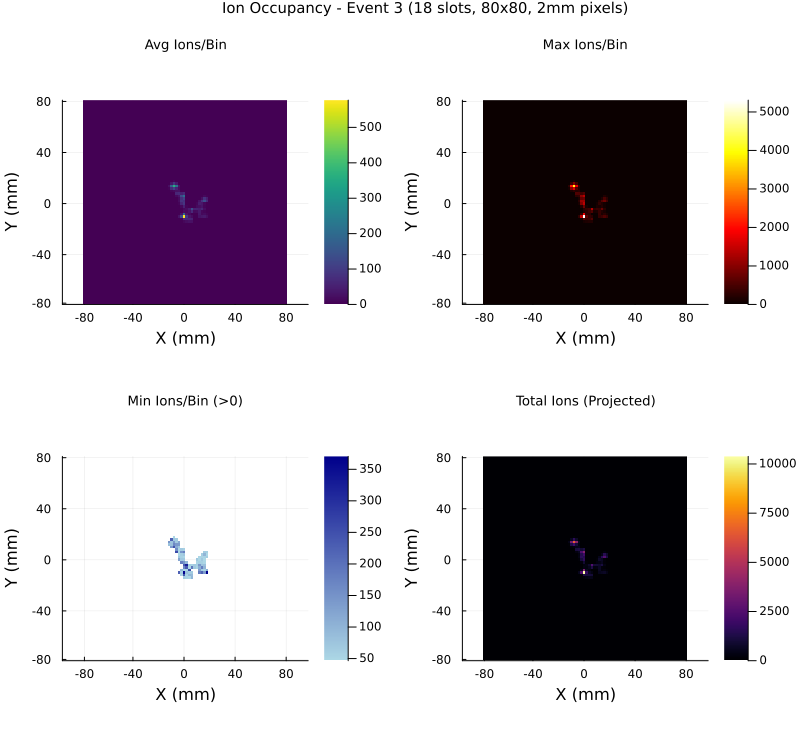}
    \caption{Typical occupancy heatmaps for \bbonu\ events, showing the spatial distribution of ion hits across the pixel plane for representative simulated decays.}
    \label{fig:oc2d}
\end{figure}
As shown in Figure~\ref{fig:sigmatvsl} the maximum diffusion of the ion track over the full drift is approximately \SI{2}{mm}. On average (tracks produced in the middle of the drift) the diffusion is of the order of \SI{1}{mm}. 

We also need the characteristic extension of the ion track
in the transverse plane (which fits within a region of approximately
$16 \times 16~\mathrm{cm^2}$), its extension along the drift direction
($\sim 100$~mm), and the ion drift velocity
($\sim 100$~mm/s at 15~bar and a drift field of 200~V/cm). 

The key requirements for the detector, then, are:
\begin{enumerate}
\item {\bf Containment of the track in XY}, requires an active area of $\mathrm{160 \times 160\ mm^2}$ 
\item{\bf Sample of the track along Z}, requires sampling the track every \SI{5}{ms} over \SI{1}{s}
\item{\bf Resolution along XY}, depends on diffusion and pitch ($p$) between pixels. Choosing $\mathrm{p = 2\ mm}$,  provides a resolution of
$ \sigma_{\mathrm{p}} = \frac{p}{\sqrt{12}} \simeq 0.58~\mathrm{mm}$, which conveniently oversamples diffusion. 
\item{\bf Resolution along Z}, sampling every \SI{5}{ms}, provides a Z resolution of 
\SI{0.5}{mm}, which oversamples diffusion. 
\item{\bf Photon-track separation} is defined by the pitch. A minimum separation of 
\SI{2}{mm} maximises the compromise between acceptable signal efficiency and background rejection, given the ion diffusion. 
\end{enumerate}

\subsection{Direct ion imaging with CMOS Topmetal sensors}

Direct collection of drifting ions on a segmented charge-sensitive plane, as proposed in 
\cite{Nygren_2018, mei2020topmetalcmosdirectcharge}
 is a natural solution in this regime. CMOS Topmetal sensors provide exposed metal electrodes directly connected to per-pixel charge-sensitive amplifiers (CSAs), enabling long integration times and frame-based readout well matched to slow ion drift.

The Topmetal II sensor~\cite{An2016TopmetalII} established the basic architecture for direct
charge collection using CMOS technology. Each pixel integrates an exposed metal pad, a CSA
with femtofarad-scale feedback capacitance, and a frame-based readout scheme. This work
demonstrated that sub-100~$e^{-}$ equivalent noise charge can be achieved for direct charge
sensing, providing the technological foundation for subsequent developments.

The use of Topmetal sensors to measure both the
topology and the energy of ion tracks produced in the double beta decay of \SE\ in a
proposed SeF$_6$ \bbonu\ experiment,  was studied in \cite{Nygren_2018}. The key insight was the use of (relatively) small collection electrodes
combined with electrostatic focusing structures to achieve high collection efficiency while
maintaining very low input capacitance. Stringent noise requirements
($\sim 30$--$50~e^{-}$ ENC) were necessary in order to preserve calorimetric information when
summing charge over a large number of pixels.

This approach was further investigated in 
\cite{mei2020topmetalcmosdirectcharge}, where the Topmetal-S design --- with an emphasis on
system-level integration --- was presented. Small CMOS dies operated at relatively large pitch
($\sim 8$~mm) and on-chip digitisation were employed in order to enable tiling over very
large areas. This choice undersampled the transverse diffusion of the ions trading spatial resolution (and gamma-track separation) in exchange for a manageable channel count and system complexity.

The common goal of both \cite{Nygren_2018} and \cite{mei2020topmetalcmosdirectcharge} was to extract detailed topology and an energy resolution at the level of $\sim 1\%$ FWHM from the ion signal, while fully instrumenting large detection planes.

For a ton-scale HPXe detector with diameter
$D \sim 3.2~\mathrm{m}$, a fully instrumented ion-imaging plane would instrument an active area of approximately $8~\mathrm{m^2}$. With a pitch of $p = 2$~mm, the total number of pixels would exceed 2 million. This is a serious challenge in terms of manufacturing, cost, channel density, power consumption, and integration. 

\subsection{Return to ITACA}

ITACA operates in a fundamentally different regime. The MARS system (section~\ref{sec:mars}) eliminates the need to instrument a full detection plane simultaneously. Instead, the 
instrumented area is limited to $16 \times 16~\mathrm{cm^2}$.
Keeping our design pitch of \SI{2}{mm} one needs
$80 \times 80 = 6400$ pixels, a factor ${\sim}300$ less than for a fully instrumented CMOS plane.

In addition, ITACA is an electroluminescent HPXe detector capable of achieving sub-percent
energy resolution through measurement of the electron track. Obtaining a comparable energy
resolution from the ion signal alone would require a drastic reduction in pixel count and
would significantly complicate the electronics. For this reason, energy reconstruction from
ions is not a driving design goal of the CMOS ion-imaging plane.

\subsection{The NAUSICA concept}

\begin{figure}[!htbp]
    \centering
    \includegraphics[scale=0.80]{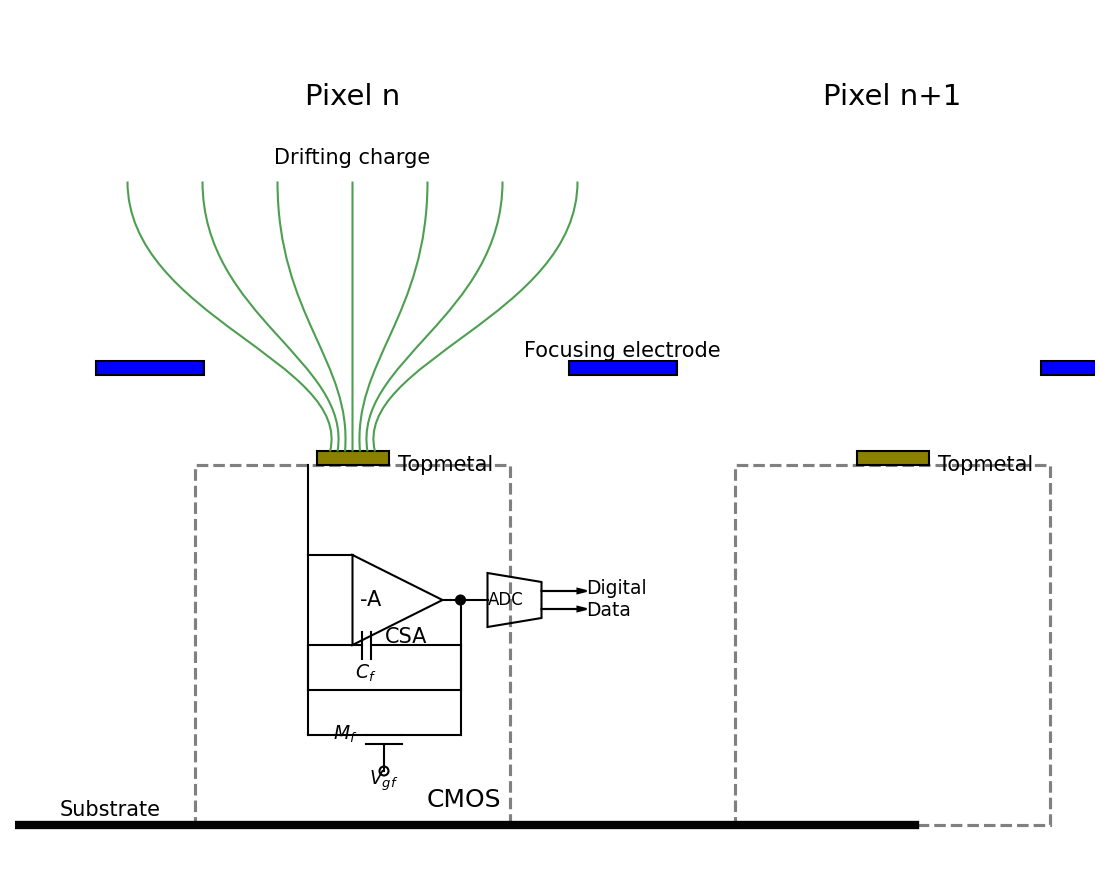}
    \caption{Sketch of the NAUSICA concept (adapted from \cite{Nygren_2018}). An bare electrode in the topmost layer of CMOS (Topmetal, shown as a yellow rectangle) collects charge in gas directly. Internal
Charge Sensitive Amplifier (CSA) converts charge signal into voltage waveform, which is then digitized by the integrated ADC (located in the periphery). Digital signals are transmitted oﬀ the chip. An elevated and separately biased electrode (blue rectangles) creates a focusing electrostatic field between itself and the Topmetal electrode (green curves).}
    \label{fig:nausicaa}
\end{figure}

Consequently, we adopt a deliberately simplified Topmetal-based sensor for ITACA, named
\textbf{NAUSICA} (Non-Amplified Ultra-Slow Ion CMOS ASIC), implemented in a 180~nm mixed-signal CMOS technology (Figure~\ref{fig:nausicaa}) with the following specifications.

\subsubsection*{Charge collection pads and focusing}

Each pixel includes an exposed metal charge-collection pad with dimensions
$    A_{\mathrm{pad}} = 0.5 \times 0.5~\mathrm{mm^2}.$
The pad size is chosen to minimise input
capacitance and electronic noise. Efficient charge collection across the full pixel cell
is ensured by electrostatic focusing electrodes.

\subsubsection*{Readout mode and temporal sampling}

Ion charge collection is continuous in time. The longitudinal coordinate is reconstructed
from time sampling, using the known ion drift velocity. To achieve a longitudinal resolution
of order 1~mm, the sensor is read out in frames with a period
$
    \Delta t = 10~\mathrm{ms},
$
corresponding to a frame rate of 100~Hz. The full ion track is therefore sampled over
approximately 100 frames.

\subsubsection*{Pixel-level analogue front-end}

Each pixel implements a minimal analogue front-end optimised for slow signals:
\begin{itemize}
    \item An exposed metal pad directly connected to the input of a charge-sensitive
    amplifier (CSA).
    \item A feedback capacitor with nominal value
    $C_f \simeq 5~\mathrm{fF}$.
    \item A MOS pseudo-resistor operating in the subthreshold regime to provide continuous
    discharge.
\end{itemize}

The feedback network defines an integration time constant
\begin{equation}
    \tau_f = R_f C_f \sim 1~\mathrm{s},
\end{equation}
comparable to the full ion drift time. The value of $\tau_f$ is adjustable via a global bias
to allow operational flexibility.

For $C_f = 5~\mathrm{fF}$, a signal of $10^3~e^{-}$ corresponds to a voltage step of
approximately 30~mV, providing comfortable signal-to-noise margin for topology
reconstruction.

\subsubsection*{Noise and dynamic range}

\begin{figure}[!htbp]
    \centering
    \includegraphics[scale=0.40]{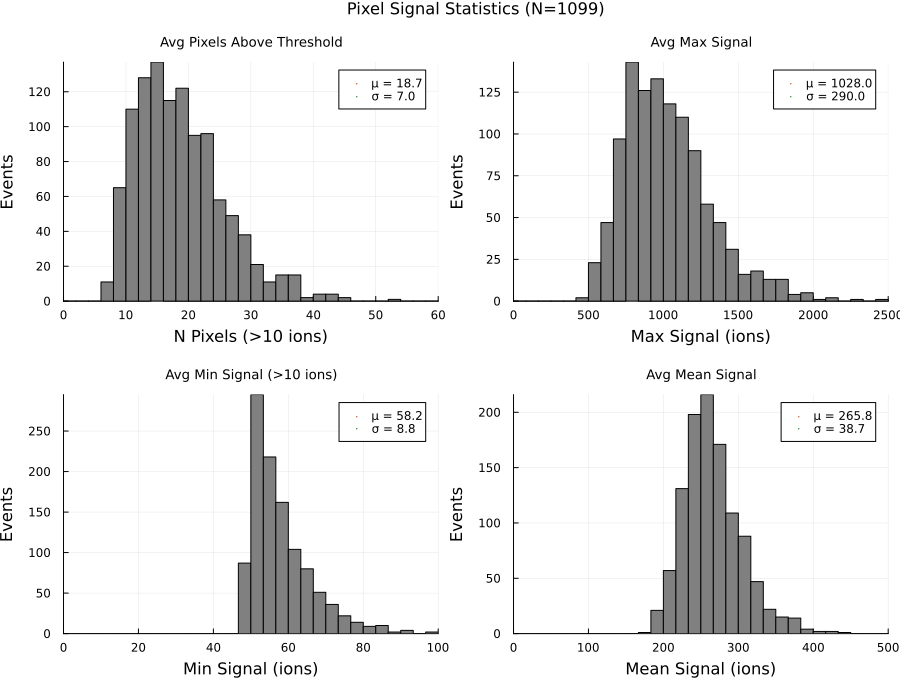}
    \caption{Top-left panel: number of pixels per frame;  top-right: average number of ions
per pixel and frame; bottom-left: minimum number of ions per pixel and frame;
bottom-right: maximum number of ions per pixel and frame.}
\label{fig:px}
\end{figure}

A \bbonu\ event (or a single electron in the ROI) produces $\sim 10^5$ ions, distributed in about
$\sim 20$ pixels per frame (Figure~\ref{fig:px}, top-left panel). The average number of ions
per pixel and frame is $\sim 260$ (Figure~\ref{fig:px}, top-right panel). The minimum signal above threshold (which we take as 10 ions) is $\sim 58$ on average and always above $\sim 45$ ions (Figure~\ref{fig:px}, bottom-left panel). The maximum signal is $\sim 1028$ ions on average and practically never exceeds 2500 (Figure~\ref{fig:px} bottom-right panel).

The target equivalent noise charge for the NAUSICA front-end is
\begin{equation}
    \mathrm{ENC} \sim 50\text{--}100~e^{-} \ \mathrm{RMS},
\end{equation}
which is sufficient to ensure robust detection of ion signals for topological purposes, since, on average, $S/N > 5$ and worst case is always above 1. 

The dynamic range is modest and amply covered with the standard CMOS ADC of 12 bits. 

\subsubsection*{Readout architecture and digitisation}

The detector is read out using a frame-based scan of all pixels. For an $80 \times 80$ pixel
array, corresponding to 6400 pixels, a frame period of 10~ms implies an aggregate sampling
rate of
$
    6.4 \times 10^5~\mathrm{samples\,s^{-1}}.
$
Digitization is implemented at the periphery of the ASIC or on an external carrier, using
one or a small number of multiplexed ADCs. A resolution of 12 bits over a full-scale range
of approximately 1~V provides quantisation noise well below the analogue noise level.

\subsubsection*{Die size and tiling strategy}

Given the large pixel pitch, only a modest number of pixels can be accommodated per die.
A baseline design adopts dies of approximately
$
    20 \times 20~\mathrm{mm^2},
$
each hosting a $10 \times 10$ pixel matrix. This results in 100 pixels per die.

A full $80 \times 80$ pixel detection plane is obtained by tiling $8 \times 8 = 64$ such dies
on a common carrier, yielding a contiguous active area of
$16 \times 16~\mathrm{cm^2}$. All periphery and I/O structures are oriented outward, ensuring
that no inactive regions are introduced within the sensitive area.

\section{Sensitivity of ITACA to \bbonu\ searches}
\label{sec:sensi}

\subsection{The effect of diffusion}
 
\begin{figure}[!htbp]
    \centering
    \includegraphics[scale=0.50]{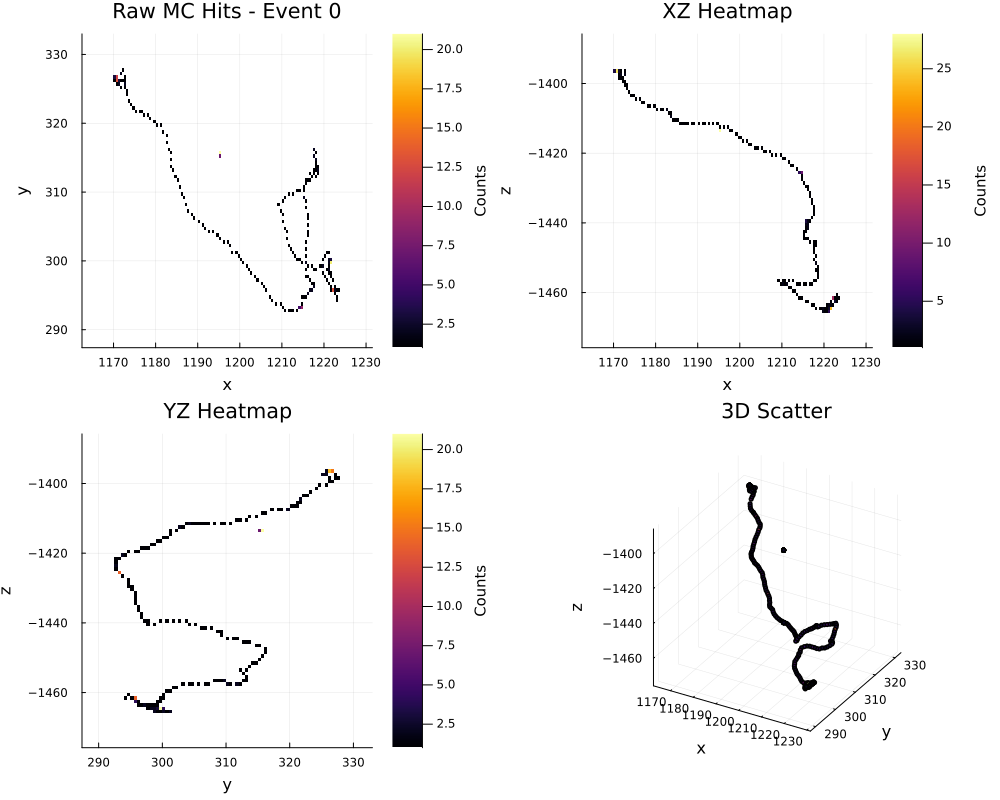}
    \caption{A simulated \bbonu\ track in HPXe.}
    \label{fig:mctrack}
\end{figure}

\begin{figure}[!htbp]
    \centering
    \includegraphics[scale=0.30]{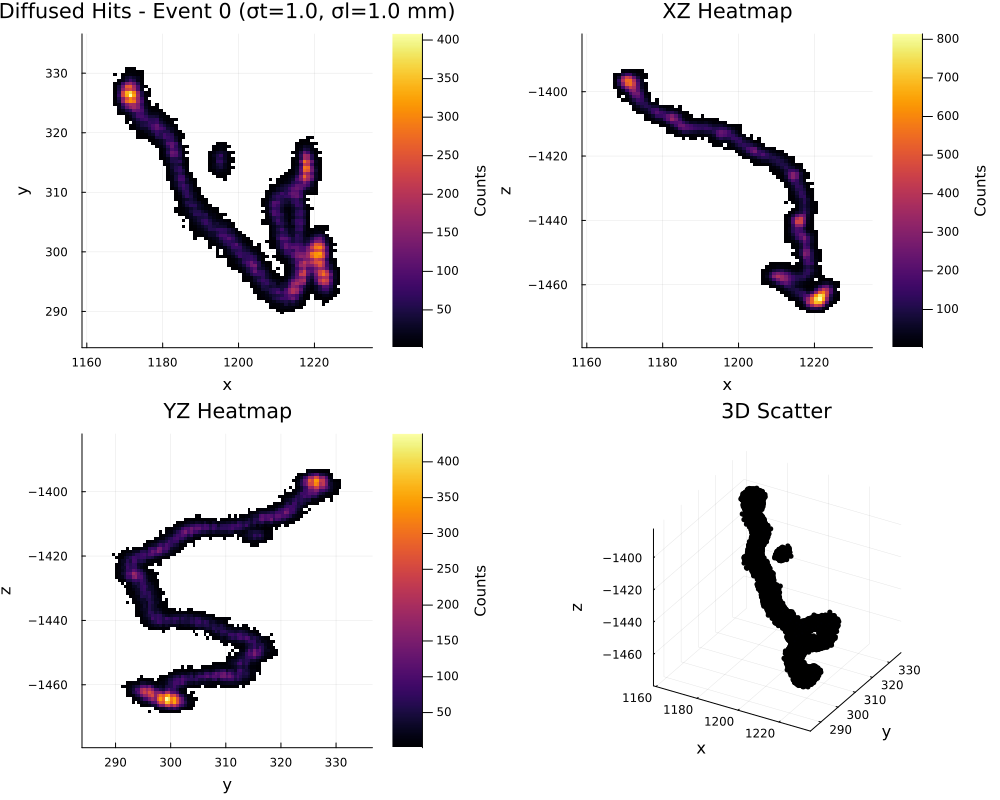}
     \includegraphics[scale=0.30]{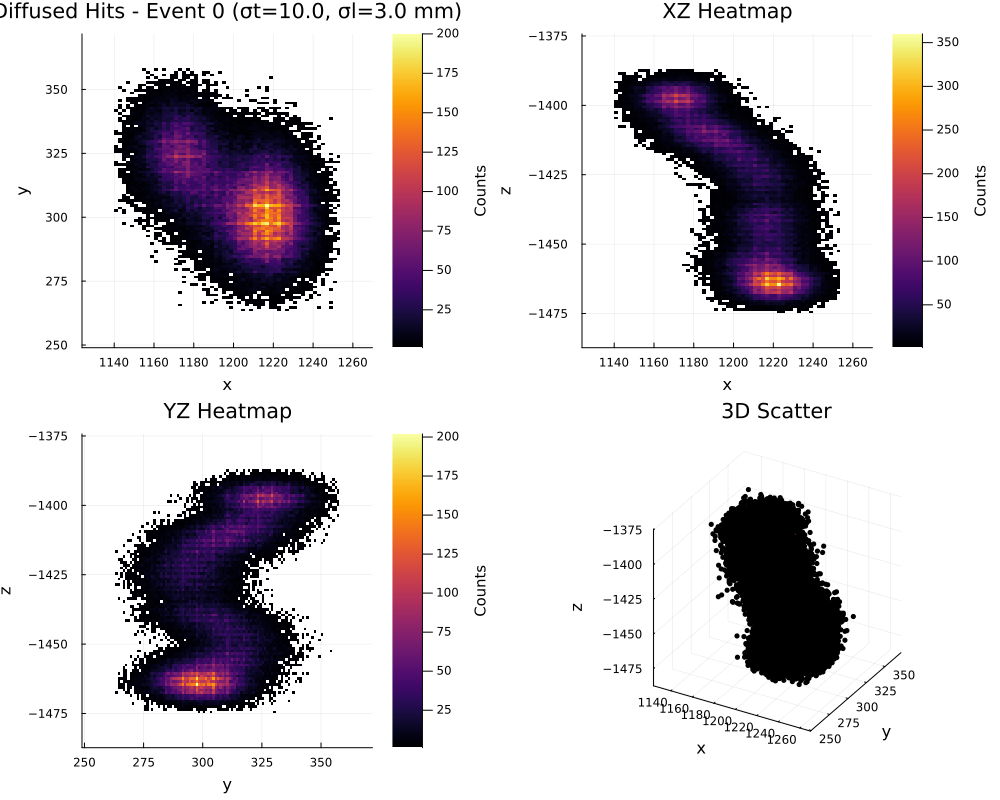}
    \caption{The effect of the diffusion in the ion track (top panel) and the electron track (bottom panel) in ITACA.}
    \label{fig:dftrack}
\end{figure}

Figure~\ref{fig:mctrack} displays the ionisation pattern from two electrons generated in a simulated \bbonu\ event. Multiple scattering bends their paths, which end in dense ``blobs" of energy deposition at the Bragg peaks. 

As shown in Figure~\ref{fig:sigmatvsl}, diffusion affects differently the ion and the electron track.
In the case of the electron track, in addition to larger diffusion one has to add the spatial spread of electroluminescent light as electrons are accelerated in the EL region. 
Figure~\ref{fig:dftrack} shows the ion track (top panel) and the electron track (bottom panel) for the Monte Carlo event depicted in Figure~\ref{fig:mctrack}.  

To separate signal and background events in the search for \bbonu\ decays, a HPXe TPC exploits the different topology of the two electrons emitted in a double beta decay, and the single electrons (often accompanied by additional energy depositions) which constitute the main backgrounds. The three main sources of background are:  the $\gamma$ line of 2447 keV arising from \BI\ decays, ($\gamma_{\BI}$); the $\gamma$ line of 2615 keV arising from \TL\ decays ($\gamma_{\TL}$), and the single electrons from the \Xe{137} $\beta$ decay with energies compatible with the ROI of the \bbonu\ search (typically 1 FWHM around \Qbb).  Given that the ROI width for EL detectors is of the order of \SIrange{15}{25}{keV}, while the distance between \Qbb\ and the $\gamma_{\BI}$ photopeak is \SI{10}{keV}, topology becomes especially important to suppress \BI\ background source, but is also needed to reduce the flat background from \Xe{137} decays and Compton interactions of the $\gamma_{\TL}$.

The $\gamma_{\BI}$ photopeak produces a single electron, accompanied 85\% of the time by the characteristic de-excitation X-ray of xenon (with an energy of about 30 keV), and occasionally Bremsstrahlung emission.  The  $\gamma_{\TL}$~introduces background in the ROI via Compton electrons, which come together with the scattered Compton photon and eventually Bremsstrahlung. The \Xe{137} $\beta$ decay produces a single electron, with end point energy of 4.17 MeV, thus intersecting the ROI. This electron can also emit Bremsstrahlung. 

In all cases, therefore, imposing a single track with no extra energy deposition is advantageous to separate signal from backgrounds. Furthermore, an HPXe TPC is able to separate single electrons from double electrons produced in a \bbonu\ decay. The signature of the double beta decay is a single connected track with two energy blobs of similar intensity in each of the track extremes, while single electrons only have one energetic blob at the end of the trajectory. 

Reducing diffusion has two positive effects. On one side, it permits a better separation between ``satellite'' energy deposits (such as the 30 keV X-ray signalling a $\gamma_{\BI}$ photoelectric interaction) and the main electron track. On the other, it allows resolving better the two blobs of a signal event. Both effects can be seen in Figure~\ref{fig:dftrack}. 

\subsection{Monte Carlo simulations}

To quantify the expected sensitivity of ITACA we have written a GEANT-4 \cite{GEANT4:2002zbu} based simulation of the detector which includes the main contributions of the radiogenic backgrounds.

The detector geometry consists of:
\begin{itemize}
 \item \textbf{ICS}: The Inner Copper Shield, modelled as a copper shell of  \SI{3}{cm} and two copper disks of \SI{3}{cm}. The true thickness of the ICS (\SI{15}{cm}) is included using a calculation that corrects for the self-shielded residual background. The ICS shields the fiducial region defined by the TPC and the BFD (along the barrel) and the TPC, the anode instrumentation and the cathode instrumentation (along the axis), from the radioactivity of the pressure vessel, made of ultra-pure titanium, which is fully attenuated by the \SI{15}{cm} of copper.
 \item \textbf{BFD}: The Barrel Fibre Detector, modelled as a PTFE shell, \SI{5}{mm} thick, located between the ICS and the TPC, along the barrel. 
 \item \textbf{TPC}: The Time Projection Chamber itself: a cylinder filled with high pressure xenon gas with $\text{R} = \SI{160}{cm}$ and $\text{L} = \SI{150}{cm}$.
\item \textbf{FC rings}: The Field Cage, modelled as 10 copper thin rings of \SI{159.4}{cm} mean radius and \SI{1.2}{cm\squared} cross-section.
 \item \textbf{SiPM boards}: These are Kapton boards holding SiPM arrays. They are modelled as a thin disk of radius  \SI{160}{cm}, \SI{0.2}{mm} thick.
\item \textbf{Light-Guides}: The light guides are modelled as a HDPE honeycomb structure of radius  \SI{160}{cm}, \SI{9}{cm}  thick.
\item \textbf{FAT-GEM}: modelled as a HDPE disk with holes of radius  \SI{160}{cm},  \SI{5}{mm} thick, with \SI{6}{mm}  holes at \SI{10}{mm} pitch.
\item \textbf{Cathode}: Fe316Ti wire mesh (200~\(\mu\)m wire, \SI{5}{mm} pitch)
\item \textbf{MARS}: All the main components are simulated, including Ti longerons, Kapton skin/cables, HDPE ribs, etc.
\end{itemize}

\begin{table}[htbp]
\centering
\caption{Specific activities of materials used in the ITACA detector.
         Values are taken from \cite{NEXT:2020amj, Alvarez:2012as}.}
\label{tab:specific_activities}
\begin{tabular}{lrr}
\toprule
Material & \textsuperscript{214}Bi (Bq/kg) & \textsuperscript{208}Tl (Bq/kg) \\
\midrule
Cu & $1.20 \times 10^{-6}$ & $1.40 \times 10^{-6}$ \\
PTFE & $1.00 \times 10^{-7}$ & $1.28 \times 10^{-6}$ \\
HDPE & $6.20 \times 10^{-6}$ & $8.00 \times 10^{-6}$ \\
Fe316Ti & $1.90 \times 10^{-3}$ & $4.00 \times 10^{-4}$ \\
Ti & $9.30 \times 10^{-4}$ & $2.20 \times 10^{-4}$ \\
Kapton & $8.00 \times 10^{-5}$ & $1.10 \times 10^{-4}$ \\
\bottomrule
\end{tabular}
\end{table}

\begin{table}[htbp]
\centering
\caption{ITACA detector radioactivity budget. Activities are computed from
         material masses and specific activities for \textsuperscript{214}Bi and
         \textsuperscript{208}Tl. The ICS (Inner Copper Shield) is shown for two
         configurations: \SI{3}{cm}  inner layer only, and full \SI{15}{cm}  with self-shielding
         correction for the outer \SI{12}{cm}.}
\label{tab:itaca_activity}
\begin{tabular}{llrrr}
\toprule
Component & Material & Mass (kg) & \textsuperscript{214}Bi (mBq) & \textsuperscript{208}Tl (mBq) \\
\midrule
BFD & PTFE & 171.2 & 0.0171 & 0.2191 \\
ICS (3cm) & Cu & 9208.2 & 11.05 & 12.89 \\
ICS (15 cm) & Cu &  13699 & 16.44 & 19.18 \\
FC rings & Cu & 107.7 & 0.1292 & 0.1508 \\
Light Guides & HDPE & 268.6 & 1.67 & 2.15 \\
FAT GEM & HDPE & 26.00 & 0.1612 & 0.2080 \\
SiPM boards & Kapton & 2.28 & 0.1827 & 0.2512 \\
Cathode & Fe316Ti & 0.7954 & 1.51 & 0.3182 \\
MARS (Ti) & Ti & 0.3006 & 0.2796 & 0.0661 \\
MARS (Kapton) & Kapton & 0.0511 & $4.09 \times 10^{-3}$ & $5.62 \times 10^{-3}$ \\
MARS (HDPE) & HDPE & 0.3458 & $2.14 \times 10^{-3}$ & $2.77 \times 10^{-3}$ \\
\end{tabular}
\end{table}

Table~\ref{tab:specific_activities} lists the specific activities of all materials
used in the ITACA detector radioactivity budget calculations \cite{NEXT:2020amj, Alvarez:2012as}.
Table~\ref{tab:itaca_activity} shows the full radioactivity budget of the ITACA detector.  The ICS is shown for two configurations: \SI{3}{cm} thickness (used in the Monte Carlo calculations), and the  full \SI{15}{cm} thickness, including the self-shielding correction for the outer \SI{12}{cm}. The leading source of background is the ICS, even when using very radiopure copper, due to the large surface exposed to copper residual radioactivity and the high density of copper (which, on the other hand, helps to self-shield a large part of the radiation coming from the ICS).

\subsection{Analysis}

\begin{table}[htbp]
\centering
\caption{ITACA Efficiencies for Bi-214 (Ion and Electron Tracks)}
\label{tab:efficiency_bi214_combined}
\begin{tabular}{lrrrrr}
\toprule
Component & $\varepsilon_{\mathrm{geom}}$ & $\varepsilon_{\mathrm{ecut}}$ & $\varepsilon_{\mathrm{fid}}$ & $\varepsilon_{\mathrm{1trk}}^{\mathrm{ion}}$ & $\varepsilon_{\mathrm{1trk}}^{\mathrm{ele}}$ \\
\midrule
BFD & $1.29 \times 10^{-3}$ & 0.7016 & $9.04 \times 10^{-4}$ & 0.0006 & 0.0020 \\
ICS (15 cm) & $2.93 \times 10^{-4}$ & 0.7842 & $2.30 \times 10^{-4}$ & 0.0113 & 0.0357 \\
FC rings & $2.93 \times 10^{-4}$ & 0.7842 & $2.30 \times 10^{-4}$ & 0.0113 & 0.0357 \\
Cathode/Anode regions & $1.27 \times 10^{-3}$ & 0.7042 & $8.91 \times 10^{-4}$ & 0.0006 & 0.0012 \\
\bottomrule
\end{tabular}
\end{table}

\begin{table}[htbp]
\centering
\caption{ITACA Efficiencies for Tl-208 (Ion and Electron Tracks)}
\label{tab:efficiency_tl208_combined}
\begin{tabular}{lrrrrr}
\toprule
Component & $\varepsilon_{\mathrm{geom}}$ & $\varepsilon_{\mathrm{ecut}}$ & $\varepsilon_{\mathrm{fid}}$ & $\varepsilon_{\mathrm{1trk}}^{\mathrm{ion}}$ & $\varepsilon_{\mathrm{1trk}}^{\mathrm{ele}}$ \\
\midrule
BFD & $4.81 \times 10^{-2}$ & 0.1540 & $7.41 \times 10^{-3}$ & 0.0012 & 0.0056 \\
ICS (15 cm) & $1.88 \times 10^{-2}$ & 0.1205 & $2.26 \times 10^{-3}$ & 0.0024 & 0.0140 \\
FC rings & $1.88 \times 10^{-2}$ & 0.1205 & $2.26 \times 10^{-3}$ & 0.0024 & 0.0140 \\
Cathode/Anode regions & $4.66 \times 10^{-2}$ & 0.1521 & $7.08 \times 10^{-3}$ & 0.0010 & 0.0066 \\
\bottomrule
\end{tabular}
\end{table}

\begin{table}[htbp]
\centering
\caption{ITACA Signal Efficiency for $0\nu\beta\beta$ (Ion and Electron Tracks)}
\label{tab:efficiency_0nubb_combined}
\begin{tabular}{lrrrrr}
\toprule
 & $\varepsilon_{\mathrm{geom}}$ & $\varepsilon_{\mathrm{ecut}}$ & $\varepsilon_{\mathrm{fid}}$ & $\varepsilon_{\mathrm{1trk}}^{\mathrm{ion}}$ & $\varepsilon_{\mathrm{1trk}}^{\mathrm{ele}}$ \\
\midrule
$0\nu\beta\beta$ & 0.9500 & 0.9878 & 0.9384 & 0.5750 & 0.7630 \\
\midrule
\textbf{Total signal efficiency} & & & & \textbf{0.5396} & \textbf{0.7160} \\
\bottomrule
\end{tabular}
\end{table}

 The analysis performs, first, a simple ``fiducial cut'', namely the events must be contained in the fiducial volume of the TPC and have an energy in the range \SIrange{2400}{2500}{keV}. We define $\varepsilon_{\mathrm{geom}}$ as the geometry cut efficiency (events contained in the fiducial region over the total number of events generated in a given region); 
 $\varepsilon_{\mathrm{ecut}}$ as the ``fiducial energy cut" efficiency (events in the range
 \SIrange{2400}{2500}{keV} over total). Finally, we define
 $\varepsilon_{\mathrm{fid}} = \varepsilon_{\mathrm{geom}} \times \varepsilon_{\mathrm{ecut}}$.
  Notice that the combined fiducial cut already suppresses the radioactive background by some three orders of magnitude, while keeping almost all the signal events which are mostly contained in the detector (since the \bbonu\ decays occur in the gas), and have an energy near \Qbb. 
  
 The next and crucial cut is the ``single track'' requirement. A \bbonu\ candidate is defined as a single, connected track, with no extra energy depositions in the fiducial volume. As discussed above, this cut is a powerful tool to reduce backgrounds, which have more satellite energy 
 depositions than the signal, at a relatively modest cost for the signal efficiency. Reconstructing the ion track is advantageous since the smaller diffusion and pitch allows to separate better the main track from the satellites. Somewhat counterintuitively, this results in a slightly worse efficiency for the signal when reconstructed with the ion track instead of the electron track. This happens because the reconstruction of the electron track misses very often bremsstrahlung photons nearby the main track, while the ion track sees them clearly and thus rejects the event.
 
 Tables~\ref{tab:efficiency_bi214_combined}, \ref{tab:efficiency_tl208_combined} and
\ref{tab:efficiency_0nubb_combined} show the selection efficiencies for \BI\ and \TL\ backgrounds generated in the various regions of the detector, as well as for the signal. Notice that, in all cases, the rejection factor of the single track cut is higher for the ion track than for the electron track. 

\begin{table}[htbp]
\centering
\caption{ITACA Bi-214 Background (Ion and Electron Tracks)}
\label{tab:background_bi214_combined}
\begin{tabular}{lrr}
\toprule
Component & Ion (ev/yr) & Electron (ev/yr) \\
\midrule
BFD & $2.93 \times 10^{-4}$ & $9.76 \times 10^{-4}$ \\
ICS (15 cm) & 1.35 & 4.25 \\
FC rings & $1.06 \times 10^{-2}$ & $3.34 \times 10^{-2}$ \\
Light Guides & $2.82 \times 10^{-2}$ & $5.63 \times 10^{-2}$ \\
FAT GEM & $2.72 \times 10^{-3}$ & $5.44 \times 10^{-3}$ \\
SiPM boards & $3.08 \times 10^{-3}$ & $6.16 \times 10^{-3}$ \\
Cathode & $2.55 \times 10^{-2}$ & $5.09 \times 10^{-2}$ \\
MARS Ti & $4.72 \times 10^{-3}$ & $9.43 \times 10^{-3}$ \\
MARS Kapton & $6.90 \times 10^{-5}$ & $1.38 \times 10^{-4}$ \\
MARS HDPE & $3.61 \times 10^{-5}$ & $7.22 \times 10^{-5}$ \\
\midrule
\textbf{TOTAL} & \textbf{1.42} & \textbf{4.42} \\
\bottomrule
\end{tabular}
\end{table}

\begin{table}[htbp]
\centering
\caption{ITACA Tl-208 Background (Ion and Electron Tracks)}
\label{tab:background_tl208_combined}
\begin{tabular}{lrr}
\toprule
Component & Ion (ev/yr) & Electron (ev/yr) \\
\midrule
BFD & $6.15 \times 10^{-2}$ & 0.287 \\
ICS (15 cm) & 3.28 & 19.2 \\
FC rings & $2.58 \times 10^{-2}$ & 0.151 \\
Light Guides & 0.481 & 3.17 \\
FAT GEM & $4.65 \times 10^{-2}$ & 0.307 \\
SiPM boards & $5.62 \times 10^{-2}$ & 0.371 \\
Cathode & $7.11 \times 10^{-2}$ & 0.469 \\
MARS Ti & $1.48 \times 10^{-2}$ & $9.75 \times 10^{-2}$ \\
MARS Kapton & $1.26 \times 10^{-3}$ & $8.29 \times 10^{-3}$ \\
MARS HDPE & $6.19 \times 10^{-4}$ & $4.09 \times 10^{-3}$ \\
\midrule
\textbf{TOTAL} & \textbf{4.04} & \textbf{24.0} \\
\bottomrule
\end{tabular}
\end{table}

\begin{table}[htbp]
\centering
\caption{ITACA Total Background Summary (Ion and Electron Tracks)}
\label{tab:background_total_combined}
\begin{tabular}{lrrr}
\toprule
Isotope & Ion (ev/yr) & Electron (ev/yr) & Ratio \\
\midrule
Bi-214 & 1.42 & 4.42 & 3.11 \\
Tl-208 & 4.04 & 24.0 & 5.94 \\
\midrule
\textbf{Total} & \textbf{5.46} & \textbf{28.4} & \textbf{5.21} \\
\bottomrule
\end{tabular}
\end{table}

Tables~\ref{tab:background_bi214_combined}, \ref{tab:background_tl208_combined} and
\ref{tab:background_total_combined} show the total number of events per year passing the cuts imposed so far, for \BI, \TL\ and both combined. The measurement of the ion track
affords a combined rejection factor of 5.2. 

The next step is to use the topology to distinguish between signal events, which are two connected electrons ending their trajectories in ``two blobs'' (as in Figure~\ref{fig:dftrack}) from single electron background events, originated in the chamber and thus with a single blob. 

\begin{figure}[!htbp]
    \centering
    \includegraphics[scale=0.80]{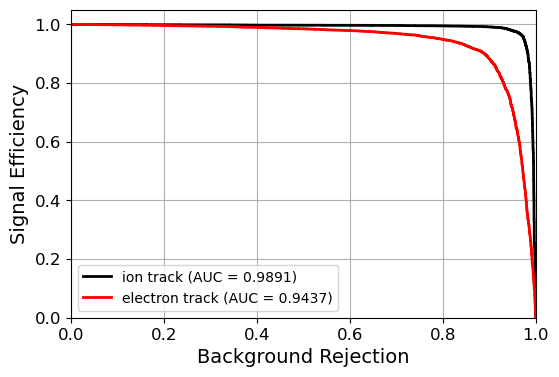}
    \caption{ROC curves describing the separation between single electrons (from \BI) and double electrons (from \bbonu\ events). The black line corresponds to the ion track and the red line to the electron track.}
    \label{fig:roc}
\end{figure}

While it is possible to formulate this condition as a cut on the energy of the blobs, much better results are obtained by using a convolutional neural network that is trained to separate signal and background events, as amply discussed in \cite{Kekic2021}.

Here, again, the ion track provides a better separation than the electron track, since the reduced diffusion allows better discrimination. The separation between single electrons (from \BI\ or \TL, which give identical curves) and double electrons (from \bbonu\ events) is shown in the ROC curves corresponding to the ion (black line) and electron (red line) tracks in Figure~\ref{fig:roc}. Fixing 80\% efficiency for the signal, the background fraction from \BI\ (\TL) is 
$10^{-2}$\ for the ion track and $7 \times 10^{-2}$ for the electron track. Thus, the use of the ion track provides an extra factor of seven in rejection power. The combination of the two selections result in an improvement of rejection power of a factor $\sim 35$. 

Indeed, the expected background after the topology selection in the case of the ion track is 
$\sim 0.05$ events per ton and year, in the very wide ROI (\SIrange{2400}{2500}{keV}) selected so far, to be compared with $\sim 2$ events per ton and year.


The remaining selection criterion is a cut restricting the energy of the events to a narrower ROI. This is needed, anyway, to suppress \bbtnu\ events, as well as the only relevant cosmogenic background in a HPXe TPC (the $\beta$ decays of \XEO\, which we discuss briefly later). 
Assuming a resolution $\sim 0.7$\% FWHM, which appears feasible after the recent results of NEXT-100~\cite{NEXT:2025ozn} and the AXEL experiment~\cite{Yoshida:2020mut},
and a symmetric ROI,
the number of background events passing the cuts is 75\% for \BI\ and 
30.7\% for  \TL. The net background rate for the conventional, pure xenon HPXeEL is
$\sim 0.7$ events per ton and year. 
In the case of ITACA, the corresponding number is 0.02 events.




Notice that a conventional HPXeEL still has considerable margin to improve, even without the ion track. First, a Xe/He mixture (at 0.9/0.1) can be used instead of pure xenon. This possibility, under investigation by the NEXT collaboration~\cite{NEXT:2019oxh} reduces the transverse diffusion by a factor 2.2 (to about \SI{4}{mm} at 15 bar after \SI{1}{m} drift) and gains roughly a factor 4 in overall rejection power, bringing the background rate of the HPXeEL (with Xe/He) to 0.3 (0.2) events for a resolution of $0.9$\% (0.7\%) FWHM. Second, an asymmetric ROI can purchase an extra rejection factor for \BI\ (at the expense of some signal efficiency). Overall, we find that a conventional HPXeEL with Xe/He can reach a background level in the vicinity of 0.2 events, a result consistent with the
studies carried out by the NEXT collaboration~\cite{NEXT:2020amj}. Still, the ability to measure the ion track allows a background rate at least an order of magnitude better. 
   
 The background arising from the $\beta$ decays of \XEO\ is of cosmogenic origin and can be effectively suppressed by operating at sufficient depth underground, as discussed in ~\cite{NEXT:2020amj}. The additional rejection power afforded by the topological signature in ITACA further helps reducing this background below that of radiogenic decays discussed here.

\section{Conclusions and Outlook}
\label{sec:conclu}
Measurement of ion tracks with a diffusion of order \SI{1}{mm} appears feasible in a HPXeEL detector. In this paper we present a conceptual design of a CMOS device needed to measure the ion track in 3D, along the lines of the \TM\ devices developed in 
\cite{An2016TopmetalII, Nygren_2018,mei2020topmetalcmosdirectcharge}. This device, that we call NAUSICA, has a finer pitch (\SI{2}{mm} between top metal pads) than those proposed by \cite{mei2020topmetalcmosdirectcharge} (\SI{8}{mm} between top metal pads). As we demonstrate in the paper, the finer pitch is essential to exploit the separation between signal and background, which includes not only the different topology between single electrons (produced by the radiogenic backgrounds as well as the decays of \XEO) and signal double-electrons, but also a larger number of satellite photons near the track in the case of backgrounds. To exploit this latter feature the pitch must be of the same order than the diffusion. This choice, however, implies a large number of channels per unit surface, as well as a tighter integration than the sensor developed in \cite{mei2020topmetalcmosdirectcharge}, but it appears well within the existing technology.

We have also presented a conceptual design of the MARS system, which 
exploits the long delay between electron signal in the anode and ion arrival to the cathode to position the NAUSICA sensor at the arrival point. The positioning ---including the sweep of the arm, radial movement of the carrier, lift of the ion plate and alignment--- takes about \SI{1}{s}, implying an efficiency of about 93\%. The analysis of gas disturbances shows that lifting the ion plate by \SI{10}{mm} ensures that ions are not affected by the gas disturbances created by the arm motion. The long-term disturbance of the gas is blocked by the IFG, a structure with a dual role: blocking the gas, and focusing the incoming ions into the pads of the CMOS sensor. 

MARS is a complex device, but uses well understood, standard mechanical solutions. The key benefit of the system is that it allows scaling the ITACA concept to large detector diameters without scaling the cost and complexity of NAUSICA. For the 1 ton detector, covering the full cathode with CMOS sensors would require two million channels, to be compared with 6,400 channels required by the small ion detector proposed here. The ratio between both configurations exceeds 300. Conceivably the technology can be extended to a large detector of diameter \SI{5}{m} and length \SI{2.5}{m} holding \SI{4.2}{\tonne} of xenon. Such a detector would require 5 million channels and the ratio with respect to the MARS sensor would approach three orders of magnitude.
It follows that the tradeoff between mechanical and electronic complexity (including also the need to digitise and extract a huge number of signals and dissipate the associated heat) favours the former. 

We have carried out a full Monte Carlo study, comparing the performance of the ITACA detector (an HPXeEL able to measure the ion track) with a conventional HPXeEL based on pure xenon. We find that a factor $\sim 30$ larger background rejection can be achieved. Under the assumptions about background budget presented in \cite{NEXT:2020amj}, we obtain a total background rate of 0.02 counts per ton per year in a symmetric, 1 FWHM ROI around \Qbb. This tiny background rate affords, in particular, considerable leverage over the very strict background budget considered in \cite{NEXT:2020amj}.  

ITACA is an HPXeEL TPC building on the technologies developed for the NEXT detectors over the last fifteen years As such, it offers several unique advantages, including the ability to fiducialise the event and excellent energy resolution. At the same time, the measurement of the ion track solves the main drawback of  electroluminescence, namely the blurring due to large diffusion in the electron track. 

The detector design presented here deploys a mass of one ton and is conceived as a compromise between mass and complexity. As mentioned above, larger detectors are conceivable without scaling the number of electronic channels needed to read the ions. The combination of potentially very large exposures and tiny background levels will make ITACA an ideal tool to reach sensitivities to the \bbonu\ decay in excess of $10^{28}$~yr, required to probe the normal neutrino-mass hierarchy and beyond.

\FloatBarrier
\section*{Acknowledgments}

Discussions with our colleagues in the NEXT collaboration and the BOLD R\&D program are warmly acknowledged.


\bibliographystyle{pool/JHEP}
\bibliography{pool/NextRefs}

\end{document}